\documentclass[journal,10pt]{IEEEtran}
\usepackage{cite}
\usepackage{amsmath,amssymb,amsfonts}
\usepackage{algorithmic}
\usepackage{graphicx,color}
\usepackage{textcomp}
\usepackage{xcolor}
\usepackage{algorithm,algorithmic}
\def\BibTeX{{\rm B\kern-.05em{\sc i\kern-.025em b}\kern-.08em
    T\kern-.1667em\lower.7ex\hbox{E}\kern-.125emX}}
\AtBeginDocument{\definecolor{tmlcncolor}{cmyk}{0.93,0.59,0.15,0.02}\definecolor{NavyBlue}{RGB}{0,86,125}}

\def\OJlogo{\vspace{-4pt}\includegraphics[height=18pt]{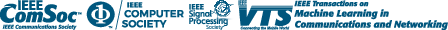}}
\def\seclogo{\vspace{10pt}\includegraphics[height=18pt]{new_logo}}

\usepackage{soul}
\usepackage{subfig}
\usepackage{multirow}  
\usepackage[nolist]{acronym}
    
 \usepackage{bm}

\newcommand{\figref}[1]{Fig.~{\ref{#1}}}

\newcommand{\compl}{\mathbb{C}}         % complex number field, e.g., x \in \compl
\newcommand{\ma}  [1]{ \bm{#1} } % matrix (upper case) and vector (lower case)
\newcommand{\Ex}[1]{\mathrm{E}\left[ #1\right]} % expectation
\newcommand{\set} [1]{{\mathcal {#1}}} % define set
\newcommand{\Kon} {\set{K}_{\text{on}}} % K_on set
\newcommand{\Kp} {\set{K}_{\text{p}}} % K_on set
\newcommand{\Kd} {\set{K}_{\text{d}}} % K_on set
\newcommand{\Kn} {\set{K}_{\text{n}}} % K_on set
\begin{acronym}
\acro{DSRC}{dedicated short-range communications}
\acro{C-ITS}{cooperative intelligent transport system}
\acro{RSU}{road side unit}
\acro{Bi}{bi-directional}
\acro{TR}{throughput}
\acro{TDL}{tapped delay line}
\acro{SRNN}{Simple RNN}
\acro{ITS}{Intelligent Transportation Systems}
\acro{IEEE}{Institute of Electrical and Electronics Engineers}
\acro{WAVE}{Wireless Access in Vehicular Environment}
\acro{V2V}{vehicle-to-vehicle}
\acro{V2I}{vehicle-to-infrastructure}
\acro{CCH}{control channel}
\acro{SCH}{service channels}
\acro{STS}{short training symbols}
\acro{LTS}{long training symbols}
\acro{SS}{signal symbol}
\acro{SoA}{state-of-the-art}
\acro{DPA}{data-pilot aided}
\acro{STA}{spectral temporal averaging}
\acro{EL}{ensemble learning}
\acro{FL}{frame length}
\acro{CDP}{constructed data pilots}
\acro{TRFI}{time-domain reliable test frequency domain interpolation}
\acro{MMSE-VP}{minimum mean square error using virtual pilots}
\acro{iCDP}{improved CDP}
\acro{SBS}{symbol-by-symbol}
\acro{FBF}{frame-by-frame}
\acro{E-TRFI}{Enhanced TRFI}
\acro{SR-CNN}{super resolution CNN}
\acro{DN-CNN}{denoising CNN}
\acro{RBF}{radial basis function}
\acro{CNN}{convolutional neural network}
\acro{TS-ChannelNet}{temporal spectral ChannelNet}
\acro{WSSUS}{wide-sense stationary uncorrelated scattering}
\acro{TDR}{transmission data rate}
\acro{LSTM}{long short-term memory}
\acro{ALS}{accurate LS}
\acro{SLS}{simple LS}
\acro{ChannelNet}{channel network}
\acro{ADD-TT}{average decision-directed with time truncation}
\acro{WI}{weighted interpolation}
\acro{DD}{decision-directed}
\acro{SR-ConvLSTM}{super resolution convolutional long short-term memory}
\acro{RS}{reliable subcarriers}
\acro{URS}{unreliable subcarriers}
\acro{AE-DNN}{auto-encoder deep neural network}
\acro{AE}{auto-encoder}
\acro{T-DFT}{truncated discrete Fourier transform}
\acro{TA-TDFT}{temporal averaging T-DFT}
\acro{TA}{time averaging}
\acro{PDP}{power delay profile}
\acro{1G}{first generation}
\acro{2G}{second generation}
\acro{3G}{third generation}
\acro{3GPP}{Third Generation Partnership Project}
\acro{4G}{fourth generation}
\acro{5G}{fifth generation}
\acro{802.11}{IEEE 802.11 specifications}
\acro{A/D}{analog-to-digital}
\acro{ADC}{analog-to-digital}
\acro{AM}{amplitude modulation}
\acro{AP}{access point}
\acro{AR}{augmented reality}
\acro{ASIC}{application-specific integrated circuit}
\acro{ASIP}{Application Specific Integrated Processors}
\acro{AWGN}{additive white Gaussian noise}
\acro{BCJR}{Bahl, Cocke, Jelinek and Raviv}
\acro{BER}{bit error rate}
\acro{BFDM}{bi-orthogonal frequency division multiplexing}
\acro{BPSK}{binary phase shift keying}
\acro{BS}{base stations}
\acro{CA}{carrier aggregation}
\acro{CAF}{cyclic autocorrelation function}
\acro{Car-2-x}{car-to-car and car-to-infrastructure communication}
\acro{CAZAC}{constant amplitude zero autocorrelation waveform}
\acro{CB-FMT}{cyclic block filtered multitone}
\acro{CCDF}{complementary cumulative density function}
\acro{CDF}{cumulative density function}
\acro{CDMA}{code-division multiple access}
\acro{CFO}{carrier frequency offset}
\acro{CIR}{channel impulse response}
\acro{CM}{complex multiplication}
\acro{COFDM}{coded-\acs{OFDM}}
\acro{CoMP}{coordinated multi point}
\acro{COQAM}{cyclic OQAM}
\acro{CP}{cyclic prefix}
\acro{CR}{cognitive radio}
\acro{CRC}{cyclic redundancy check}
\acro{CRLB}{Cram\'{e}r-Rao lower bound}
\acro{CS}{cyclic suffix}
\acro{CSI}{channel state information}
\acro{CSMA}{carrier-sense multiple access}
\acro{CWCU}{component-wise conditionally unbiased}
\acro{D/A}{digital-to-analog}
\acro{D2D}{device-to-device}
\acro{DAC}{digital-to-analog}
\acro{DC}{direct current}
\acro{DFE}{decision feedback equalizer}
\acro{DFT}{discrete Fourier transform}
\acro{DL}{deep learning}
\acro{DMT}{discrete multitone}
\acro{DNN}{deep neural network}
\acro{FNN}{feed-forward neural network}
\acro{DSA}{dynamic spectrum access}
\acro{DSL}{digital subscriber line}
\acro{DSP}{digital signal processor}
\acro{DTFT}{discrete-time Fourier transform}
\acro{DVB}{digital video broadcasting}
\acro{DVB-T}{terrestrial digital video broadcasting}
\acro{DWMT}{discrete wavelet multi tone}
\acro{DZT}{discrete Zak transform}
\acro{E2E}{end-to-end}
\acro{eNodeB}{evolved node b base station}
\acro{E-SNR}{effective signal-to-noise ratio}
\acro{EVD}{eigenvalue decomposition}
\acro{FBMC}{filter bank multicarrier}
\acro{FD}{frequency-domain}
\acro{FDD}{frequency-division duplexing}
\acro{FDE}{frequency domain equalization}
\acro{FDM}{frequency division multiplex}
\acro{FDMA}{frequency-division multiple access}
\acro{FEC}{forward error correction}
\acro{FER}{frame error rate}
\acro{FFT}{fast Fourier transform}
\acro{FIR}{finite impulse response}
\acro{FM}		{frequency modulation}
\acro{FMT}{filtered multi tone}
\acro{FO}{frequency offset}
\acro{F-OFDM}{filtered-\acs{OFDM}}
\acro{FPGA}{field programmable gate array}
\acro{FSC}{frequency selective channel}
\acro{FS-OQAM-GFDM}{frequency-shift OQAM-GFDM}
\acro{FT}{Fourier transform}
\acro{FTD}{fractional time delay}
\acro{FTN}{faster-than-Nyquist signaling}
\acro{GFDM}{generalized frequency division multiplexing}
\acro{GFDMA}{generalized frequency division multiple access}
\acro{GMC-CDM}{generalized	multicarrier code-division multiplexing}
\acro{GNSS}{global navigation satellite system}
\acro{GS}{guard symbols}
\acro{GSM}{Groupe Sp\'{e}cial Mobile}
\acro{GUI}{graphical user interface}
\acro{H2H}{human-to-human}
\acro{H2M}{human-to-machine}
\acro{HTC}{human type communication}
\acro{I}{in-phase}
\acro{i.i.d.}{independent and identically distributed}
\acro{IB}{in-band}
\acro{IBI}{inter-block interference}
\acro{IC}{interference cancellation}
\acro{ICI}{inter-carrier interference}
\acro{ICT}{information and communication technologies}
\acro{ICV}{information coefficient vector}
\acro{IDFT}{inverse discrete Fourier transform}
\acro{IDMA}{interleave division multiple access}
\acro{IEEE}{institute of electrical and electronics engineers}
\acro{IF}{intermediate frequency}
\acro{IFFT}{inverse fast Fourier transform}
\acro{IoT}{Internet of Things}
\acro{IOTA}{isotropic orthogonal transform algorithm}
\acro{IP}{internet protocole}
\acro{IP-core}{intellectual property core}
\acro{ISDB-T}{terrestrial integrated services digital broadcasting}
\acro{ISDN}{integrated services digital network}
\acro{ISI}{inter-symbol interference}
\acro{ITU}{International Telecommunication Union}
\acro{IUI}{inter-user interference}
\acro{LAN}{local area netwrok}
\acro{LLR}{log-likelihood ratio}
\acro{LMMSE}{linear minimum mean square error}
\acro{LNA}{low noise amplifier}
\acro{LO}{local oscillator}
\acro{LOS}{line-of-sight}
\acro{LP}{low-pass}
\acro{LPF}{low-pass filter}
\acro{LS}{least squares}
\acro{LTE}{Long Term Evolution}
\acro{LTE-A}{LTE-Advanced}
\acro{LTIV}{linear time invariant}
\acro{LTV}{linear time-variant}
\acro{LUT}{lookup table}
\acro{M2M}{machine-to-machine}
\acro{MA}{multiple access}
\acro{MAC}{multiple access control}
\acro{MAP}{maximum a posteriori}
\acro{MC}{multicarrier}
\acro{MCA}{multicarrier access}
\acro{MCM}{multicarrier modulation}
\acro{MCS}{modulation coding scheme}
\acro{MF}{matched filter}
\acro{MF-SIC}{matched filter with successive interference cancellation}
\acro{MIMO}{multiple-input, multiple-output}
\acro{MISO}{multiple-input single-output}
\acro{ML}{machien learning}
\acro{MLD}{maximum likelihood detection}
\acro{MLE}{maximum likelihood estimator}
\acro{MMSE}{minimum mean squared error}
\acro{MRC}{maximum ratio combining}
\acro{MS}{mobile stations}
\acro{MSE}{mean squared error}
\acro{MSK}{Minimum-shift keying}
\acro{MSSS}[MSSS]	{mean-square signal separation}
\acro{MTC}{machine type communication}
\acro{MU}{multi user}
\acro{MVUE}{minimum variance unbiased estimator}
\acro{NEF}{noise enhancement factor}
\acro{NLOS}{non-line-of-sight}
\acro{NMSE}{normalized mean-squared error}
\acro{NOMA}{non-orthogonal multiple access}
\acro{NPR}{near-perfect reconstruction}
\acro{NRZ}{non-return-to-zero}
\acro{OFDM}{orthogonal frequency division multiplexing}
\acro{OFDMA}{orthogonal frequency division multiple access}
\acro{OOB}{out-of-band}
\acro{OQAM}{offset quadrature amplitude modulation}
\acro{OQPSK}{offset quadrature phase shift keying}
\acro{OTFS}{orthogonal time frequency space}
\acro{PA}{power amplifier}
\acro{PAM}{pulse amplitude modulation}
\acro{PAPR}{peak-to-average power ratio}
\acro{PC-CC}{parallel concatenated convolutional code}
\acro{PCP}{pseudo-circular pre/post-amble}
\acro{PD}{probability of detection}
\acro{pdf}{probability density function}
\acro{PDF}{probability distribution function}

\acro{PFA}{probability of false alarm}
\acro{PHY}{physical layer}
\acro{PIC}{parallel interference cancellation}
\acro{PLC}{power line communication}
\acro{PMF}{probability mass function}
\acro{PN}{pseudo noise}
\acro{ppm}{parts per million}
\acro{PRB}{physical resource block}
\acro{PRB}{physical resource block}
\acro{PSD}{power spectral density}
\acro{Q}{quadrature-phase}
\acro{QAM}{quadrature amplitude modulation}
\acro{QoS}{quality of service}
\acro{QPSK}{quadrature phase shift keying}
\acro{R/W}{read-or-write}
\acro{RAM}{random-access memmory}
\acro{RAN}{radio access network}
\acro{RAT}{radio access technologies}
\acro{RC}{raised cosine}
\acro{RF}{radio frequency}
\acro{rms}{root mean square}
\acro{RRC}{root raised cosine}
\acro{RW}{read-and-write}
\acro{SC}{single-carrier}
\acro{SCA}{single-carrier access}
\acro{SC-FDE}{single-carrier with frequency domain equalization}
\acro{SC-FDM}{single-carrier frequency division multiplexing}
\acro{SC-FDMA}{single-carrier frequency division multiple access}
\acro{SD}{sphere decoding}
\acro{SDD}{space-division duplexing}
\acro{SDMA}{space division multiple access}
\acro{SDR}{software-defined radio}
\acro{SDW}{software-defined waveform}
\acro{SEFDM}{spectrally efficient frequency division multiplexing}
\acro{SE-FDM}{spectrally efficient frequency division multiplexing}
\acro{SER}{symbol error rate}
\acro{SIC}{successive interference cancellation}
\acro{SINR}{signal-to-interference-plus-noise ratio}
\acro{SIR}{signal-to-interference ratio}
\acro{SISO}{single-input, single-output}
\acro{SMS}{Short Message Service}
\acro{SNR}{signal-to-noise ratio}
\acro{STC}{space-time coding}
\acro{STFT}{short-time Fourier transform}
\acro{STO}{symbol time offset}
\acro{SU}{single user}
\acro{SVD}{singular value decomposition}
\acro{TD}{time-domain}	
\acro{TDD}{time-division duplexing}
\acro{TDMA}{time-division multiple access}
\acro{TFL}{time-frequency localization}
\acro{TO}{time offset}
\acro{TS-OQAM-GFDM}{time-shifted OQAM-GFDM}
\acro{UE}{user equipment}
\acro{UFMC}{universally filtered multicarrier}
\acro{UL}{uplink}
\acro{US}{uncorrelated scattering}
\acro{USB}{universal serial bus}
\acro{UW}{unique word}
\acro{VLC}{visible light communications}
\acro{VR}{virtual reality}
\acro{WCP}{windowing and \acs{CP}}	\acro{ANN}{artificial neural network}
\acro{GRU}{gated recurrent unit}
\acro{RNN}{recurrent neural network}
\acro{WHT}{Walsh-Hadamard transform}
\acro{WiMAX}{worldwide interoperability for microwave access}
\acro{WLAN}{wireless local area network}
\acro{W-OFDM}{windowed-\acs{OFDM}}	
\acro{WOLA}{windowing and overlapping}	
\acro{WSS}{wide-sense stationary}
\acro{ZCT}{Zadoff-Chu transform}
\acro{ZF}{zero-forcing}
\acro{ZMCSCG}{zero-mean circularly-symmetric complex Gaussian}
\acro{ZP}{zero-padding}
\acro{ZT}{zero-tail}
\end{acronym}

\usepackage{pgfplots}
\usepackage{tikz}
\usetikzlibrary{patterns}
\usepackage{pdfpages}
\pgfplotsset{compat=1.17}
\begin{document}

\title{RNN Based Channel Estimation in Doubly Selective Environments}

\author{Abdul~Karim~Gizzini~\IEEEmembership{Member,~IEEE},~and~Marwa~Chafii~\IEEEmembership{Member,~IEEE}

\thanks{

Abdul Karim Gizzini is with ETIS, UMR8051, CY Cergy Paris Université, ENSEA, CNRS, France (e-mail: abdulkarim.gizzini@ensea.fr).

Marwa Chafii is with the Engineering Division, New York University (NYU) Abu Dhabi, 129188, UAE and NYU WIRELESS, NYU Tandon School of Engineering, Brooklyn, 11201, NY (e-mail: marwa.chafii@nyu.edu).

}% <-this % stops a space
%\thanks{Manuscript received XXX, XX, 2015; revised XXX, XX, 2015.}
}

\maketitle

\begin{abstract}
Doubly-selective channel estimation represents a key element in ensuring communication reliability in wireless systems. Due to the impact of multi-path propagation and Doppler interference in dynamic environments, doubly-selective channel estimation becomes challenging. Conventional {\ac{SBS}} and {\ac{FBF}} channel estimation schemes encounter performance degradation in high mobility scenarios due to the usage of limited training pilots. Recently, deep learning (DL) has been utilized for doubly-selective channel estimation, where {\ac{LSTM}} and {\ac{CNN}} networks are employed in the {\ac{SBS}} and {\ac{FBF}}, respectively. However, their  usage is not optimal, since LSTM suffers from long-term memory problem, whereas, CNN-based estimators require high complexity. For this purpose, we overcome these issues by proposing an optimized {\ac{RNN}}-based channel estimation schemes, where {\ac{GRU}} and Bi-GRU units are used in {\ac{SBS}} and {\ac{FBF}} channel estimation, respectively. The proposed estimators are based on the average correlation of the channel in different mobility scenarios, where several performance-complexity trade-offs are provided. Moreover, the performance of several {\ac{RNN}} networks is analyzed. The performance superiority of the proposed estimators against the recently proposed DL-based SBS and FBF estimators is demonstrated for different scenarios while recording a significant reduction in complexity. 
\end{abstract}

\section{INTRODUCTION} \label{introduction}
\IEEEPARstart{T}he recent advances in beyond 5G networks enable high data rates and low latency mobile wireless applications~\cite{ref_UAV}. Wireless communications offer mobility to different nodes within the network, however, the mobility feature has a severe negative impact on the communication reliability~\cite{ref_ICInt}.
In such environment, the wireless channel is said to be doubly-selective, i.e. varies in both time and frequency. This is due to the propagation medium, where the transmitted signals propagate through multiple paths, each having a different power, delay, and Doppler shift effect resulting from the motion of network nodes. Knowing that the accuracy of the estimated channel influences the system performance since it affects  different operations at the receiver like equalization, demodulation, and decoding. Therefore, ensuring communication reliability using accurate channel estimation is crucial, especially in high mobility scenarios~\cite{ref_Roberto_Receiver}.

In general, a few pilots are allocated within the transmitted frame in order to maintain a good transmission data rate, where the {\ac{SoA}} channel estimation schemes can be categorized into: (\textit{i}) \ac{SBS} estimators: the channel is estimated for each received symbol separately~\cite{ref_sta, ref_trfii}. (\text{ii}) \ac{FBF} estimators: where the previous, current, and future pilots are employed in the channel estimation for each received symbol \cite{ehsanfar2020uw}. The higher channel estimation accuracy can be achieved by using {\ac{FBF}} estimators, since the channel estimation of each symbol takes advantage from the knowledge of previous, current, and future allocated pilots within the frame. Unlike, {\ac{SBS}} estimators, where only the previous and current pilots are exploited in the channel estimation for each received symbol. However, the allocated pilots are insufficient for accurately tracking the doubly-selective channel. As a result, conventional \ac{SBS} channel estimation schemes use the demapped data subcarriers besides pilot subcarriers to accomplish the channel estimation task. This procedure is known as {\ac{DPA}} channel estimation, which is unreliable due to the demapping errors of the data subcarriers that are also enlarged from one symbol to another, leading to accumulated error in the channel estimation process. Moreover, the {\ac{DPA}}-based channel estimation schemes such as {\ac{STA}}~\cite{ref_sta} and {\ac{TRFI}} {\cite{ref_trfii}} are impractical solutions as they rely on many assumptions such as high correlation of the channel within the received frame. In addition, they lack robustness in highly dynamic environments. On the other hand, several 2D interpolation methods, such as {\ac{RBF}}~\cite{ref_ChannelNet} and {\ac{ADD-TT}}~\cite{ref_TS_ChannelNet} are employed in the {\ac{FBF}} channel estimation. However, the performance of these interpolation methods is limited when employed in high mobility scenarios, since they use fixed interpolation parameters. Moreover, the well-known {\ac{FBF}} estimator is the conventional 2D \ac{LMMSE} uses the channel and noise statistics in the estimation, thus, leading to comparable performance to the ideal case. However, it suffers from high complexity making it impractical in real-case scenarios. Therefore, investigating both {\ac{SBS}} and {\ac{FBF}} channel estimators with a
good trade-off complexity vs. performance is a crucial need for improving the channel estimation accuracy as well as maintaining  affordable computational complexity.

% ref_DL_Chest2,ref_DL_Chest3, ref_Wafa1,
Recently, a great success of {\ac{DL}} has been witnessed in several wireless communications applications~\cite{ ref_DL_PHY2,chafii2018enhancing}, including localization~\cite{ref_wafa2,ref_wafa3,ref_wafa4}, and channel estimation~\cite{ref_DL_Chest1,additional_ref1,additional_ref2,additional_ref3}, particularly when integrated with conventional {\ac{SBS}} and {\ac{FBF}} estimators. This success is due to the robustness, low-complexity, and good generalization ability of {\ac{DL}} algorithms making their integration into communication systems beneficial. Motivated by these advantages, {\ac{DL}} algorithms have been integrated into doubly-selective channel estimators in two different manners: (\textit{i}) \ac{FNN} and {\ac{LSTM}} networks with different architectures and configurations are employed on top of {\ac{SBS}} estimators~\cite{ref_AE_DNN,ref_STA_DNN,ref_TRFI_DNN, ref_lstm_dnn_dpa,ref_lstm_dpa_ta}. (\textit{ii}) {\acp{CNN}} are integrated into the {\ac{FBF}} estimators~\cite{ref_ChannelNet,ref_TS_ChannelNet, WI-CNN}, where the estimated channel for the whole frame is considered as a 2D low-resolution noisy image and \ac{CNN}-based processing is applied as super-resolution and denoising techniques. These \ac{SoA} {\ac{DL}}-based {\ac{SBS}} and {\ac{FBF}} still encounter a considerable performance degradation due to the poor accuracy of the employed initial channel estimation as in~\cite{ref_STA_DNN,ref_TRFI_DNN}. Moreover, they require high computational complexity due to the employed {\ac{DL}} architectures~\cite{ref_ChannelNet,ref_TS_ChannelNet,WI-CNN}.

In order to achieve better performance-complexity trade-off in different mobility scenarios according to the channel correlation, this paper sheds light on the {\ac{RNN}}-based channel estimation in doubly-selective environments for both {\ac{SBS}} and {\ac{FBF}} channel estimation, where an optimized {\ac{RNN}} networks represented by a {\ac{GRU}} and {\ac{Bi}}-{\ac{GRU}} units are used in  the proposed {\ac{SBS}} and {\ac{FBF}} channel estimators, respectively. Thus, having a low-complexity and robust channel estimation in different mobility scenarios. The proposed {\ac{GRU}}-based {\ac{SBS}} estimator uses only one {\ac{GRU}} network instead of two as the case in the recently proposed {\ac{LSTM}}-based estimator~\cite{ref_lstm_dnn_dpa}. After that, {\ac{DPA}} estimation is applied using the {\ac{GRU}} estimated channel. Finally, unlike~\cite{ref_lstm_dnn_dpa} where {\ac{FNN}} network is used for noise elimination, in the proposed {\ac{GRU}}-based estimators, {\ac{TA}} processing is employed as a noise alleviation technique where the noise alleviation ratio is calculated analytically. Moreover, motivated by the fact that {\ac{Bi}}-{\ac{RNN}} is designed to perform 2D interpolation of unknown data bounded between known data~\cite{bi-RNN}, the proposed  {\ac{Bi}}-{\ac{GRU}} channel estimator is designed to overcome the limitations of the {\ac{FBF}} {\ac{CNN}}-based channel estimation schemes, where an end-to-end 2D interpolation is performed by the proposed Bi-GRU unit. In this context,  the proposed {\ac{Bi}}-{\ac{GRU}} channel estimator employs an adaptive frame design, where comb pilot allocation is replaced by full pilot allocated symbols that are inserted periodically within the transmitted frame. As a first step, the channel is estimated at the inserted pilot symbols, after that, {\ac{Bi}}-{\ac{GRU}} acts as an end-to-end 2D interpolation unit to estimate the channel at the data symbols without the need to any initial estimation. By doing this interpolation, the proposed Bi-GRU based estimator is able to further improve  the estimation performance, unlike the {\ac{CNN}}-based estimators that work according to the noise mitigation principle~\cite{ref_SRCNN, ref_DNCNN} rather than doing actual interpolation. 
Simulation results show the performance superiority of the proposed {\ac{RNN}}-based channel estimation schemes against the {\ac{SoA}} {\ac{SBS}} and {\ac{FBF}} channel estimators while recording an outstanding computational complexity reduction. To sum up, the contributions of this paper are listed below \footnote{We would like to mention that part of this work related to the Bi-RNN based FBF channel estimation has been accepted for publication in the IEEE ICC 2023 conference~\cite{gizzini2023deep}.}:

\begin{itemize}

    \item Proposing low-complexity and robust {\ac{RNN}}-based channel estimation schemes, where an optimized {\ac{GRU}}, and {\ac{Bi}}-{\ac{GRU}} units are employed to accurately estimate the doubly-selective channel in {\ac{SBS}} and {\ac{FBF}} fashions, respectively.
    
    \item Employing {\ac{GRU}} unit as a pre-processing module to {\ac{DPA}} and {\ac{TA}} processing in {\ac{SBS}} channel estimation. Whereas an end-to-end 2D interpolation using \ac{Bi}-{\ac{GRU}} unit is proposed for {\ac{FBF}} channel estimation.

    \item Providing a brief overview of the theoretical concept of the studied {\ac{RNN}} networks.
    
    \item Analyzing the appropriate {\ac{RNN}} architectures to be employed according to the average channel correlation within the frame in different mobility scenarios, where the advantages of using the proposed optimized {\ac{GRU}} unit instead of regular {\ac{LSTM}} unit are discussed.
    
    \item Showing that the proposed {\ac{RNN}}-based channel estimators record a significant superiority over the {\ac{SoA}} {\ac{SBS}} and {\ac{FBF}} channel estimators in terms of {\ac{BER}} and throughput for different modulation orders, mobility scenarios, and frame lengths.
    
    \item Illustrating the advantage of using the {\ac{EL}} algorithm~\cite{ref_EL} in the generalization of one {\ac{DL}} model that is robust against a range of Doppler frequencies.
    
    \item Providing a detailed computational complexity analysis for the studied channel estimators, where we show that the proposed {\ac{RNN}}-based channel estimators achieve substantial reduction in complexity in comparison with the {\ac{SoA}} {\ac{SBS}} and {\ac{FBF}} channel estimators.
\end{itemize}

The remainder of this paper is organized as follows: Section~\ref{IEEE80211p_standard} presents the system model. The {\ac{SoA}} {\ac{DL}}-based channel estimation schemes are thoroughly investigated and discussed in Section~\ref{literature_review}. Section~\ref{Proposed_scheme} illustrates the framework of the proposed {\ac{RNN}}-based channel estimation schemes, besides providing a brief overview of the main {\ac{RNN}} networks integrated into the doubly-selective channel estimation. In Section~\ref{simulation_results}, different modulation orders are used to present simulation results, wherein the performance of the studied estimators is examined in terms of \ac{BER}. Detailed computational complexity analysis is provided in Section~\ref{complexity}. Finally, Section~\ref{conclusions} concludes this study. 
\section{SYSTEM MODEL} \label{IEEE80211p_standard}

Consider a frame consisting of $I$ {\ac{OFDM}} symbols.  The $i$-th transmitted frequency-domain {\ac{OFDM}} symbol  $\tilde{\ma{x}}_i[k]$, is denoted by
\begin{equation}
   \tilde{\ma{x}}_i[k] = \left\{
            \begin{array}{ll}        
                \tilde{\ma{x}}_{{i,d}}[k],&\quad k \in \Kd. \\
                \tilde{\ma{x}}_{{i,p}}[k],&\quad k \in \Kp. \\
                0,&\quad k \in \Kn. \\
            \end{array}\right.
\label{eq: xK}
\end{equation}
where $k$ refers to the subcarrier index, where $0 \leq k \leq K - 1$. Moreover, $d$ and $p$ indices refer to the transmitted data and pilot subcarriers, respectively. The total number of subcarriers is divided into $K_{\text{on}} = K_{d} + K_{p}$ subcarriers in addition to $K_{n}$ null guard band subcarriers, where 
$ \tilde{\ma{x}}_{{i,d}}[k]$ and $ \tilde{\ma{x}}_{{i,p}}[k]$ represent the modulated data symbols and the predefined pilot symbols allocated at a set of subcarriers denoted $\Kd$ and $\Kp$, respectively. The received frequency-domain {\ac{OFDM}} symbol denoted as $\tilde{\ma{y}}_{{i}}[k]$ is expressed as follows

\texorpdfstring{\vspace{-15pt}}{}

\begin{equation}
	\begin{split}
		\tilde{\ma{y}}_{{i}}[k] 
		&= \tilde{\ma{h}}_i[k] \tilde{\ma{x}}_i[k] + \tilde{\ma{v}}_i[k],~ k \in \Kon.
	\end{split}            
	\label{eq: system_model}
\end{equation}
Here, $\tilde{\ma{h}}_i[k] \in \compl^{K_{on} \times 1}$ refers to the frequency response of the doubly-selective channel at the $i$-th {\ac{OFDM}} symbol and $k$-th subcarrier. ${\tilde{\ma{v}}}_i[k]$ signifies the \ac{AWGN} of variance $\sigma^2$. As a matrix form, \eqref{eq: system_model} can be expressed as follows
 
\begin{equation}
\tilde{\ma{Y}}[k,i] = \tilde{\ma{H}}[k,i]  \tilde{\ma{X}}[k,i] + \tilde{\ma{V}}[k,i],~ k \in \Kon,
\label{eq:preamble_freq}
\end{equation}
where $\tilde{\ma{V}}[k,i] \in \compl ^{K_{\text{on}}\times I}$ and $\tilde{\ma{H}} \in \compl ^{K_{\text{on}}\times I}$ denote the {\ac{AWGN}} noise and the doubly-selective frequency response of the channel for all symbols within the transmitted \ac{OFDM} frame, respectively.
\section{SoA DL-BASED CHANNEL ESTIMATION} 
\label{literature_review}

This section presents the recently proposed {\ac{SoA}} {\ac{DL}}-based {\ac{SBS}} and {\ac{FBF}} channel estimation schemes, where the processing steps applied in each estimator are presented.

\subsection{DL-BASED SBS CHANNEL ESTIMATION SCHEMES}

In general, \ac{FNN} and {\ac{LSTM}} networks are employed in the {\ac{SBS}} channel estimation, where an optimized {\acp{FNN}} are integrated as a post-processing unit with conventional {\ac{SBS}} channel estimators as the case in the {\ac{DPA}}-{\ac{FNN}}~\cite{ref_AE_DNN}, {\ac{STA}}-{\ac{FNN}}~\cite{ref_STA_DNN}, and {\ac{TRFI}}-{\ac{FNN}}~\cite{ref_TRFI_DNN}. On the other hand, {\ac{LSTM}} networks are utilized as a pre-processing unit in the {\ac{LSTM}}-{\ac{FNN}}-{\ac{DPA}}~\cite{ref_lstm_dnn_dpa}, and {\ac{LSTM}}-{\ac{DPA}}-{\ac{TA}}~\cite{ref_lstm_dpa_ta} channel estimators.
Both implementations are helpful in improving the  accuracy of the channel estimation. However, the LSTM-based estimation illustrates a considerable superiority over the \ac{FNN}-based estimation. In this context, and since we are focusing on {\ac{RNN}}-based channel estimation, this section presents the steps applied in the {\ac{LSTM}}-based channel estimators.

\subsubsection{LSTM-FNN-DPA}

The work proposed in~\cite{ref_lstm_dnn_dpa} shows that employing the {\ac{LSTM}} processing prior to the {\ac{DPA}} estimation could lead to a significant improvement in the overall performance. In this context, two cascaded {\ac{LSTM}} and \ac{FNN} networks for both channel estimation as well as noise compensation. The LSTM-FNN-DPA estimator employs the {\ac{LS}} estimated channel at the previous and current received pilots, such that
\begin{equation}
\hat{\tilde{\ma{h}}}_{i,p}[k] = \frac{\tilde{\ma{y}}_{i}[k]}{\tilde{\ma{x}}_{p}[k]},~\hat{\tilde{\ma{h}}}_{i-1,p}[k] = \frac{\tilde{\ma{y}}_{i-1}[k]}{\tilde{\ma{x}}_{p}[k]}, ~k \in \Kp.
\label{eq:LS-pilots}
\end{equation}
The {\ac{LS}} estimated channel are fed as an input to both \ac{LSTM} and {\ac{FNN}} networks, where the \ac{LSTM}-{\ac{FNN}} estimated channel is expressed as follows
%
%  ,~ \hat{\tilde{\ma{h}}}_{\text{LSTM}_{0}}[k] = \hat{\tilde{\ma{h}}}_{\text{LS}}[k]
\begin{equation}
    \tilde{\ma{d}}_{\text{LSTM-FNN}_{i,d}}[k] =  \mathfrak{D} \big( \frac{\tilde{\ma{y}}_{i,d}[k]}{\hat{\tilde{\ma{h}}}_{\text{LSTM-FNN}_{i-1,d}}[k]}\big).
    \label{eq: proposed3}
    \end{equation}
    \begin{equation}
    \hat{\tilde{\ma{h}}}_{\text{DL}_{i,d}}[k] = \frac{\tilde{\ma{y}}_{i,d}[k]}{\ma{d}_{\text{LSTM}_{i,d}}[k]}.
    \label{eq: proposed44}
    \end{equation}
We note that, at the beginning of the frame ($i = 1$), $\hat{\tilde{\ma{h}}}_{i-1,p}[k]$ denotes the {\ac{LS}} estimated channel at the received preamble symbols, such that
    
\begin{equation}
\hat{\tilde{\ma{h}}}_{\text{LS}}[k] = \frac{\sum\limits_{u=1}^{P}\tilde{\ma{y}}_{u,p}[k]}{P\tilde{\ma{x}}_{p}[k]},~k \in \Kon.
\label{eq: LS}
\end{equation}
While this estimator can outperform the  \ac{FNN}-based estimators, it encounters a high complexity cost arising from the employment of two {\ac{DL}} networks.

\subsubsection{LSTM-DPA-TA}
The authors in~\cite{ref_lstm_dpa_ta} propose to use only an optimized LSTM network instead of two as implemented in the LSTM-FNN-DPA estimator. In addition, noise compensation is made possible by applying {\ac{TA}} processing. This methodology only requires the previous pilots $\hat{\tilde{\ma{h}}}_{i-1,p}[k]$ besides the LSTM estimated channel as an input. Then, the LSTM estimated channel is employed in the DPA estimation as follows
    \begin{equation}
    \tilde{\ma{d}}_{\text{LSTM}_{i}}[k] =  \mathfrak{D} \big( \frac{\tilde{\ma{y}}_i[k]}{\hat{\tilde{\ma{h}}}_{\text{LSTM}_{i-1}}[k]}\big)
    ,~ \hat{\tilde{\ma{h}}}_{\text{LSTM}_{0}}[k] = \hat{\tilde{\ma{h}}}_{\text{LS}}[k],
    \label{eq: proposed33}
    \end{equation}
    \begin{equation}
    \hat{\tilde{\ma{h}}}_{\text{LSTM-DPA}_{i}}[k] = \frac{\tilde{\ma{y}}_i[k]}{\tilde{\ma{d}}_{\text{LSTM}_{i}}[k]}.
    \label{eq: proposed4}
    \end{equation}

AWGN noise alleviation can be achieved by further applying {\ac{TA}} processing such that
    \begin{equation}
    \hat{\bar{\ma{h}}}_{\text{DL-TA}_{i,d}} = (1 - \frac{1}{\alpha})  \hat{\bar{\ma{h}}}_{\text{DL-TA}_{i - 1,d}} + \frac{1}{\alpha}  \hat{\bar{\ma{h}}}_{\text{LSTM-DPA}_{i,d}}.
     \label{eq: proposed5}
\end{equation}
% \hl{define what is $\alpha$}
Here, $\alpha$ denotes the utilized weighting coefficient. In~\cite{ref_lstm_dpa_ta}, the authors use a fixed $\alpha = 2$ for simplicity. Therefore, the {\ac{TA}} applied in~{\eqref{eq: proposed5}} reduces the AWGN noise power $\sigma^2$ iteratively within the received {\ac{OFDM}} frame according to the ratio
\begin{equation}
	\begin{split}
			{R}_{\text{DL-TA}_{q}} &= \left( \frac{1}{4} \right)^{(q-1)} + \sum_{j=2}^{q} \left( \frac{1}{4} \right)^{(q-j+1)}=\frac{4^{q-1} + 2}{3 \times 4^{q-1}}.
	\label{eq:noise_degradtion}
	\end{split}
\end{equation}
This corresponds to the AWGN noise power ratio of the estimated channel at the $q$-th estimated channel, where ${1 < q < I + 1}$ and ${{R}_{\text{DL-TA}_{1}} = 1}$ denotes the AWGN noise power ratio at $\hat{\tilde{\ma{h}}}_{\text{LS}}[k]$. From the derivation of ${R}_{\text{DL-TA}_{q}}$, it can be seen that the noise power decreases over the received {\ac{OFDM}} frame, i.e. the SNR increases, resulting in an overall improved performance. The full derivation of~\eqref{eq:noise_degradtion} is found in~\cite{ref_lstm_dpa_ta}.    
Even though the {\ac{LSTM}}-{\ac{DPA}}-{\ac{TA}} improves the performance compared to the {\ac{LSTM}}-{\ac{FNN}}-{\ac{DPA}} estimator, it still suffers from high computational complexity. Moreover, in Section~\ref{Proposed_scheme}, we show that employing {\ac{LSTM}} unit in the channel estimation would affect the estimation accuracy negatively. Whereas, the proposed {\ac{GRU}}-based channel estimation provides a better performance-complexity trade-off.  

\subsection{CNN-BASED FBF CHANNEL ESTIMATION SCHEMES}

In~\cite{WI-CNN}, a {\ac{CNN}} aided {\ac{WI}} channel estimation schemes have been proposed. The WI-CNN estimators use adaptive frame structure according to the mobility scenario. The idea is to avoid using comb pilot allocation and insert $Q$ pilot {\ac{OFDM}} symbols with different configurations within the transmitted {\ac{OFDM}} frame instead.  
In this context, the \ac{WI}-{\ac{CNN}} estimators employ one, two, and three pilot symbols in low, high, and very high mobility scenarios, respectively. Following the selection of the frame structure, the \ac{WI}-{\ac{CNN}} estimators proceed as follows 

\begin{itemize}
    \item \textbf{Pilot symbols channel estimation}: In order to estimate the channel at the inserted pilot symbols, the basic {\ac{LS}} denoted as \emph{{\ac{SLS}}} estimation is applied using the received preambles as shown in~\eqref{eq: LS}, and using each received pilot symbol such that
        \begin{equation}
        \hat{\tilde{\ma{h}}}_{{\text{SLS}}_{q}}[k] = \frac{\tilde{\ma{y}}^{(p)}_{q}[k]}{\tilde{\ma{p}}[k]} = \tilde{\ma{h}}_q[k] + \tilde{\ma{v}}_{{q}}[k], ~ k \in \Kon.
        \label{eq: SLSP}
        \end{equation}
        where $\tilde{\ma{v}}_{q}[k]$ represents the noise at the $q$-th received pilot symbol, $1 \leq q \leq Q$ denotes the inserted pilot symbol index, and 
        $\tilde{\ma{Y}}_{Q} = [\tilde{\ma{y}}^{(p)}_{1}, \dots,  \tilde{\ma{y}}^{(p)}_{q}, \dots, \tilde{\ma{y}}^{(p)}_{Q}] \in \compl ^{K_{\text{on}}\times Q}$.
       Moreover, \emph{{\ac{ALS}}} can be obtained by applying the {\ac{DFT}} interpolation of $\hat{\ma{h}}_{q,L}$ such that 
        \begin{equation}
        %\begin{split}
        \hat{\tilde{\ma{h}}}_{{\text{ALS}}_{q}} = \ma{F}_{\text{on}} \hat{\ma{h}}_{q,L},  ~ k \in \Kon,
        %\end{split}
        \label{eq:ALSP}
        \end{equation}
        with ${\hat{\ma{h}}}_{q,L} \in \compl ^{L\times 1}$ denotes the estimated channel impulse response at the $q$-th received pilot symbol.
         We note that the \emph{{\ac{ALS}}} and \emph{{\ac{SLS}}} are used for full pilot (FP) allocation. However, if the number of channel taps $L$ is known, the channel estimation requires only $L$ pilots in each pilot symbol, where DFT interpolation can be applied to the estimated channel impulse response  $\hat{\ma{h}}_{q,L}$ such that

         \begin{eqnarray}
    \hat{\tilde{\ma{h}}}_{\text{DFT}_{q}} = \ma{F}_{\text{on}} \hat{\ma{h}}_{q,L},  ~ k \in \Kon,
    \label{eq:dft5}
    \end{eqnarray}
    
    where $\ma{F}_{\text{on}} \in \compl^{K_{\text{on}} \times L}$ denotes the truncated DFT matrices obtained by selecting $\Kon$ rows, and $L$ columns from the $K$-DFT matrix. 
        
    \item \textbf{Data symbols channel estimation}: After estimating the channel at the inserted $Q$ pilot symbols, the WI-CNN estimator divides the received frame into several sub-frames that are grouped as follows
    \begin{equation}
          \hat{\tilde{\ma{H}}}_{q} = [\hat{\tilde{\ma{h}}}_{q-1}, \hat{\tilde{\ma{h}}}_{q}],~ q = 1, \cdots Q,
      \end{equation}
     $\hat{\tilde{\ma{h}}}_{q}$ refers to the implemented {\ac{LS}} estimation. Then, the estimated channel for the $i$-th received data {\ac{OFDM}} symbol within each sub-frame is calculated as a weighted summation of the estimated channels at the pilot symbols, such that 
      \texorpdfstring{\vspace{-2pt}}{}
      \begin{equation}
          \hat{\tilde{\ma{H}}}_{{\text{WI}}_{f}} =  \hat{\tilde{\ma{H}}}_{{f}} \ma{C}_{f},
      \label{eq:WI_LS}
      \end{equation}
      where $\hat{\tilde{\ma{H}}}_{f} \in \compl^{ K_{\text{on}} \times 2} $ denotes {\ac{LS}} estimated channels at the pilot symbols within the $f$-th sub-frame. $\ma{C}_{f} \in \mathbb{R}^{2 \times I_{f}}$ denotes the interpolation weights of the $I_{f}$ {\ac{OFDM}} data symbols within the $f$-th sub-frame. The interpolation weights of $\ma{C}_{f}$ are calculated by minimizing the {\ac{MSE}} between the ideal channel $\tilde{\ma{H}}_{{f}}$, and the {\ac{LS}} estimated channel at the {\ac{OFDM}} pilot symbols $\hat{\tilde{\ma{H}}}_{{f}}$ as derived in~\cite{ref_interpolation_matrix}. In the final step,  optimized {\ac{SR-CNN}} is employed  on top of the {\ac{WI}} estimators in a low mobility scenario, whereas an optimized {\ac{DN-CNN}} is considered in high mobility one.    
\end{itemize}
The {\ac{WI}}-{\ac{CNN}} estimators suffer from high computational complexity.  Moreover, using noise alleviation {\acp{CNN}} is not sufficient to accurately estimate the doubly-selective channel. Therefore, we propose a {\ac{Bi}}-{\ac{GRU}} channel estimator that perform 2D interpolation, unlike the {\ac{SR-CNN}} and {\ac{DN-CNN}} networks, which are based on noise alleviation techniques. As a result, performance superiority of the proposed {\ac{Bi}}-{\ac{GRU}} channel estimator can be achieved while recording a significant decrease of the computational complexity  in comparison to the {\ac{WI}}-{\ac{CNN}} estimators as illustrated in Section~\ref{simulation_results} and Section~\ref{complexity}.
\section{PROPOSED RNN-BASED CHANNEL ESTIMATION SCHEMES} \label{Proposed_scheme}

In this section, {\ac{RNN}} main concepts and extensions are first thoroughly introduced. Then, a detailed explanation of the proposed \ac{RNN} and {\ac{Bi}}-{\ac{RNN}} based schemes for {\ac{SBS}} and {\ac{FBF}} channel estimation are presented, respectively.

\subsection{RECURRENT NEURAL NETWORKS: REVIEW \label{RNN_Overv}}

\ac{RNN} is a type of \ac{ANN} designed to work with sequential data. This sequential data can be in form of time series, text, audio, video etc.
RNN uses the previous information in the sequence to produce the current output, where it is incorporated with memory to take information from prior inputs to influence the current output. This mechanism is the key essence to {\ac{RNN}} success in sequential problems. The core concept of {\acp{RNN}} is to keep/discard input data in a recurring manner. Therefore, {\ac{RNN}} gates contain Sigmoid activation. A Sigmoid activation regulates values in the range $[0,1]$, which is helpful to update or forget data because any number getting multiplied by 0 disappears and is  discarded. In contrast, any number multiplied by 1 is kept. The network can learn which data is not important, and  therefore can be forgotten or which data is important to keep. Moreover, Tanh activation is used within the {\ac{RNN}} architecture at some point in order to regulate the network in the training phase by squishing values between -1 and 1, since zero values affect negatively the training accuracy.

In order to make things clearer, the {\ac{RNN}} input is expressed in terms of the initial estimated channel at the $i$-th {\ac{OFDM}} symbol denoted by $\hat{\Bar{\ma{h}}}_{\rho_{i}} \in \mathbb{R}^{K_{\text{in}} \times 1}$, where $K_{\text{in}} = 2 K_{\text{on}}$. Since {\ac{DL}} networks work with real-valued numbers, then, 
\begin{equation}
    \hat{\Bar{\ma{h}}}_{\rho_{i}} = \text{vec} \Big\{\Re{ \big(\hat{\tilde{\ma{h}}}_{\rho_{i}}}\big) , \Im{\big(\hat{\tilde{\ma{h}}}_{\rho_{i}}}\big)\Big\}.
    \label{eq:lstm_input_stacked}
\end{equation}
Here, $\rho$ refers to the applied initial channel estimation scheme, and $\Re\{.\}$ and $\Im\{.\}$ denote the real and imaginary values of the initial estimated channel $\hat{\tilde{\ma{h}}}_{\rho_{i}} \in \compl^{K_{\text{on}} \times 1}$, respectively. Moreover, the  {\ac{RNN}} output is denoted by $\hat{\Bar{\ma{h}}}_{\rho\text{-RNN}_{i}} \in \mathbb{R}^{K_{\text{in}} \times 1}$. We note that \ac{FNN} network treats the initial estimated channels separately, where it produces the output $\hat{\Bar{\ma{h}}}_{\rho\text{-FNN}_{i}}$ for each input. By doing this single input-output mapping, the {\ac{FNN}} network is able to learn the frequency correlation of the doubly-selective channel, besides correcting the initial estimation error. On the contrary, {\ac{RNN}} network treats the initial estimated channel as a correlated sequence, where the current $\hat{\Bar{\ma{h}}}_{\rho\text{-RNN}_{i}}$ is computed using the previous {\ac{RNN}} estimated channel $\hat{\Bar{\ma{h}}}_{\rho\text{-RNN}_{i-1}}$ and the current initial  estimated channel $\hat{\Bar{\ma{h}}}_{\rho_{i}}$. This process allows the {\ac{RNN}} network to learn both frequency and time correlation of the doubly-selective channel, and thus, {\ac{RNN}} outperforms {\ac{FNN}} in the channel estimation task.

\begin{figure*}[t]
\includegraphics[width=2\columnwidth]{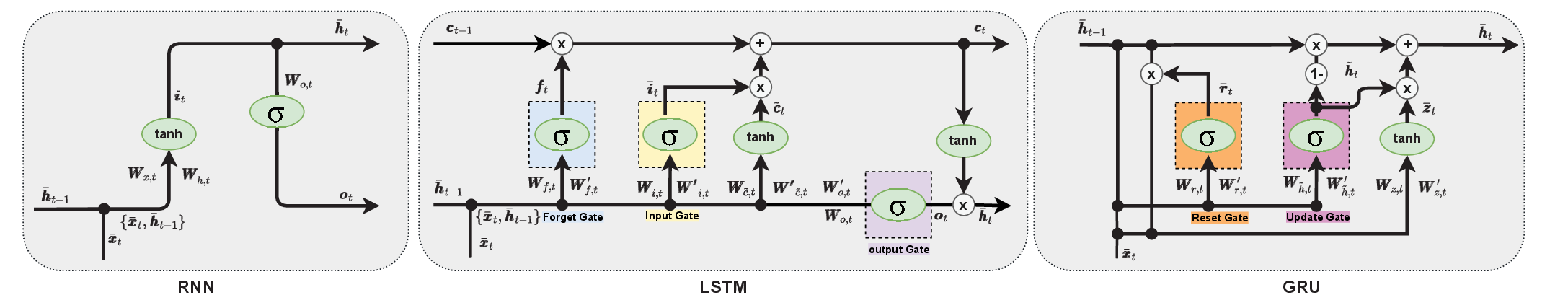}
\caption{Detailed architecture of the SRNN, LSTM, and GRU units.}
\label{fig:RNNs_architecture}
\end{figure*}

In general, there exist three main types of {\acp{RNN}}: (\textit{i}) \ac{SRNN}, (\textit{ii}) \ac{LSTM}, and (\textit{iii}) \ac{GRU}. The main difference between them is in how the input data is processed by each {\ac{RNN}} as shown in~\figref{fig:RNNs_architecture}. For simplicity, let $\hat{\Bar{\ma{h}}}_{\rho_{i}} = \bar{\ma{x}}_{t} $ and $\hat{\Bar{\ma{h}}}_{\rho\text{-RNN}_{i}} = {{{\ma{o}}}}_{t}$, where $t$ denotes the time index.
\subsubsection{{\ac{SRNN}}}
is useful when we need to look at recent information only to perform a present task, where the hidden state is constantly being rewritten in each time step. The {\ac{SRNN}} updates the current hidden state, and its output is as follows
\begin{equation}
{\bar{\ma{h}}}_{t} = \sigma (\ma{W}_{x, t}\bar{\ma{x}}_{t} + \ma{W}_{h,t}\bar{\ma{h}}_{t-1} + \bar{\ma{b}}_{h,t}),
\label{eq:RNN_hstate}
\end{equation}
\begin{equation}
{{{\ma{o}}}}_{t} = \sigma (\ma{W}_{o, t}\bar{\ma{h}}_{t} + \bar{\ma{b}}_{o,t}),
\label{eq:RNN_ostate}
\end{equation}
where ${\sigma}$ denotes the Sigmoid function, $\ma{W}_{x,t} \in \mathbb{R}^{P \times K_{in}}$,  $\ma{W}_{h,t} \in \mathbb{R}^{P \times P}$ and $\bar{\ma{b}}_{h,t} \in \mathbb{R}^{P \times 1}$ are the weight matrices and biases associated with the {\ac{SRNN}} input vector $\bar{\ma{x}}_{t} \in \mathbb{R}^{K_{in} \times 1}$ and the previous hidden state vector $\bar{\ma{h}}_{t-1} \in \mathbb{R}^{P \times 1} $, respectively. We note that at the first step, a hidden state will usually be seeded as a matrix of zeros so that it can be fed into the RNN cell together with the first input in the sequence.
Training and back-propagation in {\ac{RNN}} are similar to other forms of {\ac{ANN}}, where {\ac{RNN}} needs to be trained in order to produce accurate and desired outputs. However, when an {\ac{SRNN}} is exposed to long sequences,  it tends to lose the information because it cannot store long sequences since it focuses only on the latest output only. This problem is commonly referred to as vanishing gradients~\cite{vanish} that occurs during the training phase, where useful gradients cannot propagate from the output of the model back to the layers near the input of the model.
As a result, the {\ac{RNN}} does not learn the effect of earlier inputs and it is too difficult for RNN to preserve information over many time steps, hence, causing the short-term memory problem.
To overcome this problem, specialized versions of \ac{RNN} like {\ac{LSTM}} and {\ac{GRU}}  are created.

\subsubsection{LSTM}
is a special kind of RNN capable of learning long-term sequences.
Recall that in \ac{SRNN}, the input and hidden state from the previous time step are passed through a simple activation layer to obtain a new state.  Whereas in LSTM the process is slightly complex, where the LSTM unit  takes at each time input from three different states defined as: (\textit{i}) current input state represented by $\bar{\ma{x}}_{t}$. (\textit{ii}) short-term memory state from the previous LSTM unit denoted by $\bar{\ma{h}}_{t-1}$, and (\textit{iii}) long-term memory state from the previous LSTM unit denoted by ${\ma{c}}_{t-1}$. 
These inputs are controlled by the gates to regulate the information to be kept or discarded before passing the updated long-term and short-term information to the next LSTM unit. We can imagine these gates as filters that remove unwanted selected and irrelevant information. LSTM uses mainly three gates defined as input gate, forget gate, and output gate.

\paragraph{Forget Gate}
It decides which information from long-term memory should be kept or discarded and this is done by multiplying the incoming long-term memory by a forget vector generated by the current input and incoming short memory.  Information from the previous hidden state and information from the current input is passed through the sigmoid function. Values come out between 0 and 1. The closer to 0 means to forget, and the closer to 1 means to keep, such that
\texorpdfstring{\vspace{-5pt}}{}
\begin{equation}
{\ma{f}}_{t} = \sigma (\ma{W}_{f, t}\bar{\ma{x}}_{t} + \ma{W}^{\prime}_{f,t}\bar{\ma{h}}_{t-1} + \bar{\ma{b}}_{f,t}),
\label{eq: lstm_fg}
\end{equation}
where ${\sigma}$ denotes the Sigmoid function, $\ma{W}_{f,t} \in \mathbb{R}^{P \times K_{in}}$,  $\ma{W}^{\prime}_{f,t} \in \mathbb{R}^{P \times P}$ and $\bar{\ma{b}}_{f,t} \in \mathbb{R}^{P \times 1}$ are the forget gate weights and biases at time $t$, $\bar{\ma{x}}_{t} \in \mathbb{R}^{K_{in} \times 1}$ and $\bar{\ma{z}}_{t-1}$ represents the {\ac{LSTM}} unit input vector of size $K_{in}$, and the previous hidden state of size $P$, respectively. 
\paragraph{Input Gate}
The input gate only works with the information from the current input  $\bar{\ma{x}}_{t}$ and short-term memory $\bar{\ma{h}}_{t-1}$ from the previous step by filtering out the information that is not useful. The input gate proceeds as follows
\texorpdfstring{\vspace{-5pt}}{}
\begin{equation}
{\bar{\ma{i}}_{t}} = \sigma (\ma{W}_{\bar{\ma{i}}, t}\bar{\ma{x}}_{t} + \ma{W}^{\prime}_{\bar{\ma{i}},t}\bar{\ma{h}}_{t-1} + \bar{\ma{b}}_{\bar{\ma{i}},t}),
\label{eq:lstm_ing}
\end{equation}
\begin{equation}
{\tilde{{\ma{c}}}}_{t} = \text{tanh} (\ma{W}_{{\tilde{{\ma{c}}}}, t}\bar{\ma{x}}_{t} + \ma{W}^{\prime}_{{\tilde{{\ma{c}}}},t}\bar{\ma{h}}_{t-1} + \bar{\ma{b}}_{{\tilde{{\ma{c}}}},t}).
\label{eq: lstm_incg}
\end{equation}
Now we should have enough information to calculate the new long-term memory represented by the cell state. First, the cell state gets element-wise multiplied by the forget vector. This has a possibility of dropping values in the cell state if it gets multiplied by values near 0. Then we take the output from the input gate and do an element-wise addition that updates the cell state to new values that the neural network finds relevant. That gives us our new cell state, such that
    
\begin{equation}
{{{\ma{c}}}}_{t} = {\ma{f}}_{t} \odot {\ma{c}}_{t-1} +  \bar{\ma{i}}_{t} \odot {\tilde{{\ma{c}}}}_{t}. 
\label{eq: lstm_cell_state}
\end{equation}
Here, $\odot$ denotes the Hadamard product.

\paragraph{Output Gate}
The output gate takes the current input, the previous short-term memory, and the newly computed long-term memory~\eqref{eq: lstm_cell_state} to produce new short-term memory which will be passed on to the next time step. The output of the current time step can also be drawn from this hidden state as follows 
\begin{equation}
{\bar{{\ma{h}}}}_{t} =  {\ma{o}}_{t} \odot \text{tanh}({\ma{c}}_{t}),
\label{eq: lstm_hidden_state}
\end{equation}
\begin{equation}
{\ma{o}}_{t} = \sigma (\ma{W}_{o, t}\bar{\ma{x}}_{t} + \ma{W}^{\prime}_{o,t}\bar{\ma{h}}_{t-1} + \bar{\ma{b}}_{o,t}).
\label{eq: lstm_og}
\end{equation}
To summarize, the forget gate decides what is relevant to keep from prior steps. The input gate decides what information is relevant to add from the current step. The output gate determines what the next hidden state should be.

\subsubsection{GRU}
The GRU is the newer generation of \ac{RNN}, it is based on the same concept as the \ac{LSTM}, but with optimized architecture. The \ac{GRU} architecture gets rid of the cell state, so there is no long-term memory as the case in the {\ac{LSTM}}. Instead, the hidden state is used only to transfer information. Moreover, {\ac{GRU}} has only two gates, a reset gate, and an update gate. 

\paragraph{Reset gate}

The reset gate is used from the model to decide how much of the past information is needed to neglect. In short, it decides whether the previous hidden state is important or not, such that 
\texorpdfstring{\vspace{-8pt}}{}
\begin{equation}
{\bar{\ma{r}}}_{t} = \sigma (\ma{W}_{r, t}\bar{\ma{x}}_{t} + \ma{W}^{\prime}_{r,t}\bar{\ma{h}}_{t-1} + \bar{\ma{b}}_{r,t}),
\label{eq:gru_r}
\end{equation}
\paragraph{Update gate}
The update gate acts similarly to the forget and input gate of an LSTM. It decides what information to throw away and what new information to add from the current input. It is also responsible for determining the amount of previous information that needs to pass along to the next state, such that
\texorpdfstring{\vspace{-5pt}}{}
\begin{equation}
{\bar{\ma{z}}}_{t} = \sigma (\ma{W}_{z, t}\bar{\ma{x}}_{t} + \ma{W}^{\prime}_{z,t}\bar{\ma{h}}_{t-1} + \bar{\ma{b}}_{z,t}),
\label{eq:gru_z}
\end{equation}
\texorpdfstring{\vspace{-8pt}}{}
\begin{equation}
{\tilde{{\ma{h}}}}_{t} = \text{tanh} (\ma{W}_{{\tilde{{\ma{h}}}}, t}\bar{\ma{x}}_{t} + \ma{W}^{\prime}_{{\tilde{{\ma{h}}}},t} ({\bar{\ma{r}}}_{t} \odot \bar{\ma{h}}_{t-1}) + \bar{\ma{b}}_{{\tilde{{\ma{h}}}},t}).
\label{eq: gru2}
\end{equation}
Finally, the new hidden state is calculated as follows
    
\begin{equation}
{{{\bar{\ma{h}}}}}_{t} = (1 - {\bar{\ma{z}}}_{t}) \odot {\bar{\ma{h}}}_{t-1} +  \bar{\ma{z}}_{t} \odot {\tilde{{\ma{h}}}}_{t}. 
\label{eq:gru_new_hidden_state}
\end{equation}

To sum up, {\ac{RNN}} networks are incorporated with a memory to take advantage of the prior outputs in the prediction of the current output, and thus, the current output prediction is correlated with the previous outputs. {\ac{SRNN}} cannot store long sequences, since it only focuses  on the latest output. On the contrary, {\ac{LSTM}} is capable of learning long-term sequences, and predicting the current output is influenced by the long sequence of previous outputs. However, LSTM is not useful in all scenarios, especially, when the successive inputs become uncorrelated over time. Since predicting the current output will be affected by uncorrelated previous outputs, the prediction accuracy is negatively affected. In this context, {\ac{GRU}} provides a trade-off; it uses a short-term memory in predicting the current output, which improves the prediction accuracy in comparison with LSTM. Moreover, the GRU has two gates, whereas LSTM has three gates. Thereby, fewer training parameters and faster execution can be achieved by using GRU instead of LSTM. 
\begin{figure}[t]
\centering
\includegraphics[width=\columnwidth]{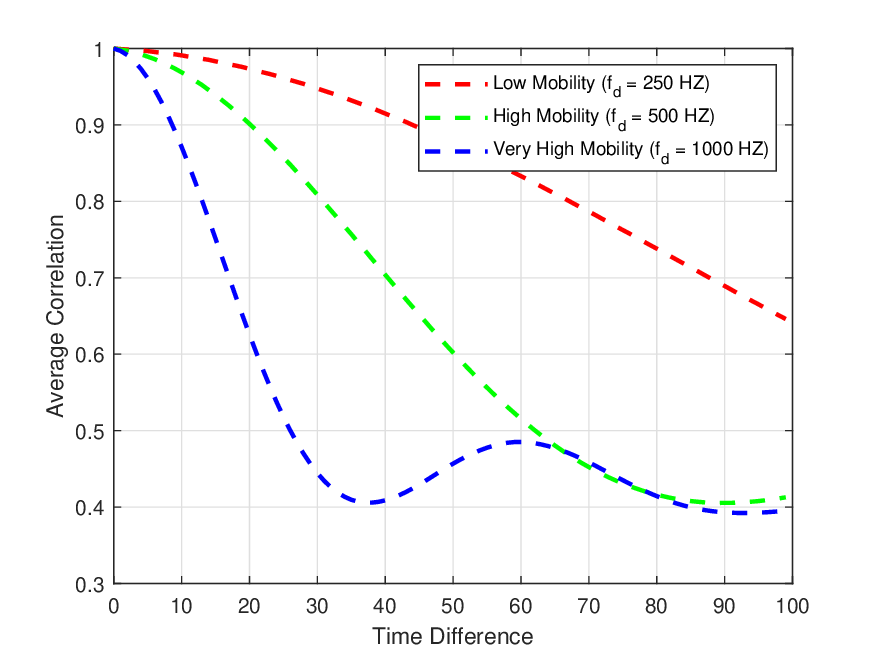}
\caption{Correlation of the channel at the first and the last OFDM symbol within the transmitted frame.}
\label{fig:channel_correlation}
\end{figure}

\begin{figure*}[t]
\includegraphics[width=2\columnwidth]{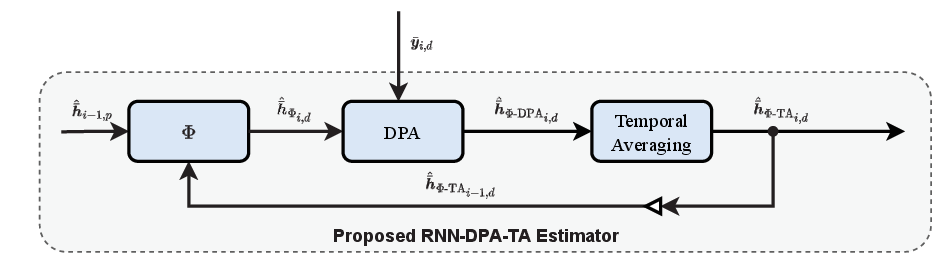}
\caption{Proposed RNN-based channel estimation schemes.}
\label{fig:Proposed_RNN_Schemes}
\end{figure*}

\subsection{PROPOSED SBS RNN-BASED CHANNEL ESTIMATOR}\label{SBS_RNN_pro}
The proposed \ac{RNN}-based estimation scheme sheds light on the ability of {\ac{GRU}} in estimating doubly-selective channels with high accuracy.
As discussed in Section~\ref{RNN_Overv},  {\ac{SRNN}} takes advantage of the previously estimated channel only while estimating the current one. Whereas, the {\ac{LSTM}} has long-term memory, which means that estimating the channel at the current {\ac{OFDM}} symbol is affected by the older estimated channels. On the other hand, {\ac{GRU}} provides a trade-off between short-term memory and complexity. Therefore, in order to decide which {\ac{RNN}} performs better in doubly-selective channel estimation, we study the average correlation between the channel at the first symbol and all successive symbols  within the transmitted {\ac{OFDM}} frame, considering the frequency-time response, such that
% \mathcolorbox{yellow}{}
\begin{equation}
\Psi_{i} = \Ex{\tilde{\ma{h}}_{1}\tilde{\ma{h}}^*_{\text{i}}},~ 2 \leq i \leq I.
\label{eq:avg_channel_correlation}
\end{equation}
Here, $\Psi_{i}$ is calculated for three mobility scenarios: (\textit{i}) Low mobility: $f_{d} = 250$ Hz, (\textit{ii}) High mobility: $f_{d} = 500$ Hz, and (\textit{iii}) Very high mobility: $f_{d} = 1000$ Hz. The detailed properties of these channel models are provided in Section~{\ref{simulation_results}}.

As shown in~\figref{fig:channel_correlation}, when the mobility increases, the average correlation $\Psi$ starts to decrease exponentially. However, as we can notice, $\Psi_{i}$ at the end of the received frame reaches around $65\%$ for low mobility scenario, while it is around $40\%$ in high and very high mobility scenarios, with a drastic decrease in the overall $\Psi_{i}$ curve in very high mobility scenarios. According to the $\Psi_{i}$ values in different mobility scenarios, we can expect that the impact of the estimated channels  at  earlier symbols would affect negatively the accuracy of the estimated channel at advanced symbols within the received OFDM frame. As a result,
we can conclude that, as the mobility increases, shorter {\ac{RNN}} memory is required in the channel estimation in order to guarantee the best possible performance. This is due to the fact that, when long {\ac{RNN}} memory is employed in a very high mobility scenario, the older estimated channels negatively impact the channel estimation at the current {\ac{OFDM}} symbol because  the estimated channels become uncorrelated, i.e. the value of $\Psi$ is low. In this context, the proposed RNN-based channel estimation scheme employs an optimized {\ac{GRU}} unit instead of {\ac{LSTM}} unit in the channel estimation process due to its shorter memory. This results in improving the accuracy of the channel estimation while recording a significant decrease in computational complexity. Moreover, we study the performance of the \ac{SRNN} unit in order to have a complete analysis of   different {\ac{RNN}} units. 

As illustrated in~\figref{fig:Proposed_RNN_Schemes}, the
{\ac{RNN}} unit is first employed to estimate the channel at the current data subcarriers, where it takes as an input the previous {\ac{LS}} estimated channels at pilot subcarriers denoted by $\hat{\Bar{\ma{h}}}_{i-1,p} \in \mathbb{R}^{2K_{p} \times 1}$, concatenated with the previously RNN-based estimated channel at the data subcarriers $\hat{\Bar{\ma{h}}}_{{\Phi\text{-TA}}_{i-1,d}} \in \mathbb{R}^{2K_{d} \times 1}$. Thus, the input and output sizes of the {\ac{RNN}} unit are $2 K_{on}$ and $2 K_{d}$, respectively. After that, the {\ac{RNN}} output is fed as an input to the {\ac{DPA}} estimation followed by {\ac{TA}} processing in order to further mitigate the impact of noise. We note that our proposed estimators consider both time and frequency selectivity, where the RNN-based pre-processing deals with the time selectivity and the DPA estimation  with the frequency selectivity. Moreover, the RNN-based estimated channel is fed as an input to the DPA estimation block, which further improves the DPA estimation accuracy. The proposed RNN-based channel estimation scheme proceeds as follows
    \begin{equation}
    \Bar{\ma{d}}_{{\Phi}_{i,d}}[k] =  \mathfrak{D} \big( \frac{\Bar{\ma{y}}_{i,d}[k]}{\hat{\Bar{\ma{h}}}_{{\Phi}_{i-1,d}}[k]}\big)
    ,~ k \in \Kd,
    \label{eq: proposed335}
    \end{equation}
    \begin{equation}
    \hat{\Bar{\ma{h}}}_{{\Phi\text{-DPA}}_{i,d}}[k] = \frac{\Bar{\ma{y}}_{i,d}[k]}{\Bar{\ma{d}}_{{\Phi}_{i,d}}[k]},
    \label{eq: proposed45}
    \end{equation}
where ${\Phi} \in \{\text{SRNN}, \text{GRU}\}$ refers to the used RNN unit, and $\hat{\Bar{\ma{h}}}_{{\Phi}_{0,d}} = \hat{\Bar{\ma{h}}}_{\text{LS}}~\forall~k \in \Kd$. Finally, to alleviate the impact of the AWGN noise, {\ac{TA}} processing is applied to the estimated channel $\hat{\Bar{\ma{h}}}_{\Phi\text{-DPA}_{i}}$ similarly as performed in~\eqref{eq: proposed5}, such that
    \begin{equation}
    \hat{\bar{\ma{h}}}_{\Phi\text{-TA}_{i,d}} = (1 - \frac{1}{\alpha})  \hat{\bar{\ma{h}}}_{\Phi\text{-TA}_{i - 1,d}} + \frac{1}{\alpha}  \hat{\bar{\ma{h}}}_{\Phi\text{-DPA}_{i,d}},
     \label{eq: proposed55}
\end{equation}
\begin{table}
	\centering
	\caption{Parameters of the proposed RNN-based channel estimation scheme.}
	\label{tb:LSTM_params}
	\begin{tabular}{l|l}
	\hline
		(SRNN units; Hidden size) & (1;48)  \\ \hline
		(GRU units; Hidden size) & (1;48)  \\ \hline
		(Bi-SRNN units; Hidden size) & (1;32)  \\ \hline
		(Bi-GRU; Hidden size) & (1;32)  \\ \hline
		Activation function              & ReLU ($y= \max(0,x)$)                     \\ \hline
		Number of epochs        & 500                                \\ \hline
		Training samples        & 16000                             \\ \hline
		Testing samples        & 2000                             \\ \hline
		Batch size          & 128                                    \\ \hline
		Optimizer       & ADAM                                       \\ \hline
		Loss function      & MSE                                     \\ \hline
		Learning rate        & 0.001                                 \\ \hline
	    Training SNR        & 40 dB                                 \\ \hline
	\end{tabular}
\end{table}
where $\alpha = 2$ for simplicity. We note that in doubly-selective channel, each two successive symbols are correlated regardless of the mobility scenario, therefore, using $\alpha = 2$ gives equal weights for the previous and current estimated channel. However, However, alpha can be fine-tuned by studying the average channel correlation between each two successive OFDM symbols, and then assigning more accurate weights to the previous and current estimated channels in~{\eqref{eq: proposed55}}.

In the proposed scheme, RNN training is performed using a high value of \ac{SNR} = 40 dB to achieve the best performance as observed in~{\cite{r20}}. The reason is that when the training is performed for low noise impact, the
RNN is able to better learn the channel correlation. In addition,  due to its good generalization ability, it can still perform well in low {\ac{SNR}} regions, where the noise is dominant. Moreover, intensive experiments are performed using the grid search algorithm~{\cite{ref_grid_search}} in order to select the best suitable RNN hyper-parameters in terms of both performance and complexity. Note that the mobility conditions can be assumed  known in most real case applications. For example, in vehicular communications, the vehicle velocity is a known parameter that
can be exchanged between all vehicular network nodes and it must be regulated according to the road conditions.  In urban environments (inside cities) the car velocity must not exceed 40 Kmphr, and thus, the model trained on low mobility can be employed. Consequently, the {\ac{RNN}} training is performed for each mobility scenario separately using the same architecture and training parameters summarized in Table~{\ref{tb:LSTM_params}}. However, when velocity information is not available, {\ac{EL}} algorithm can be used to combine the weights of several trained models so that one generalized model can be employed in all mobility scenarios as discussed in~{\ref{simulation_results}}.

\subsection{PROPOSED FBF Bi-RNN-BASED CHANNEL ESTIMATOR}

Bi-RNN networks are designed to predict unknown data that are bounded within known data~\cite{bi-RNN}. They are based on making 
the data flow through any RNN unit in both directions forward (past to future), and  backwards (future to past).  In regular RNN, the input flows in one direction, whereas, in {\ac{Bi}}-{\ac{RNN}} the input flows in both directions to get the advantage of both past and future information. By doing so, the {\ac{Bi}}-{\ac{RNN}} network will be able to predict the unknown information in the middle based on its correlation with the known past and future information.
\begin{figure*}[t]
	\centering	\includegraphics[width=2\columnwidth]{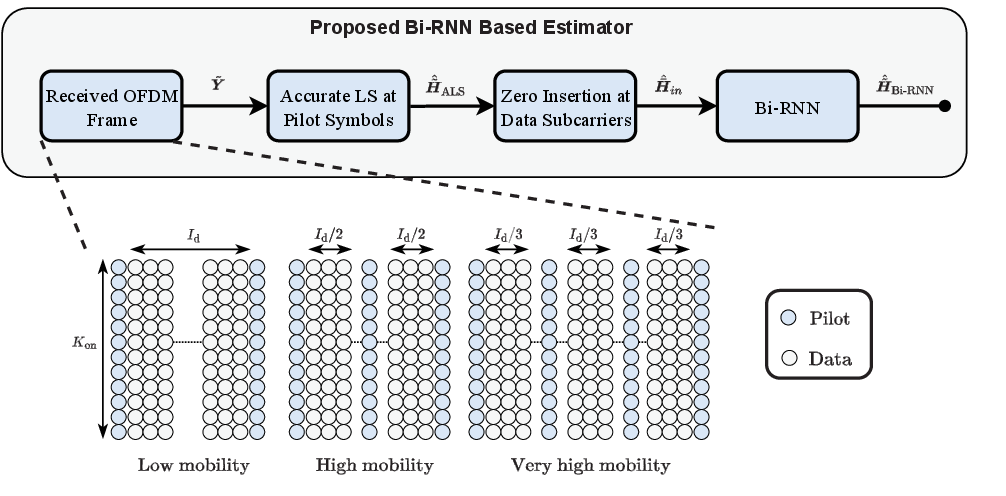}
	\caption{Proposed Bi-RNN based channel estimator block diagram.}
	\label{fig:proposed-bi-rnn}
\end{figure*}
In this context, the proposed Bi-RNN channel estimator aims to utilize the  interpolation ability of Bi-RNN networks in the {\ac{FBF}} channel estimation instead of employing high-complexity CNN networks as it is the case in the {\ac{SoA}} channel estimation schemes. The proposed Bi-RNN channel estimation scheme uses Bi-GRU unit and it inherits the adaptive frame design from the WI-CNN estimators as shown in Fig.~\ref{fig:proposed-bi-rnn}. Recall that WI-{\ac{CNN}} channel estimation performs WI interpolation at the data symbols, where CNN processing is applied to alleviate the impact of noise. However, Bi-RNNs perform 2D interpolation at the data symbols using the estimated channel at the pilot symbols without the need for any initial channel estimation at the data symbols. Thus, the proposed Bi-RNN channel estimator can be adapted to any existing protocols regardless of the pilot allocation scheme. However, the employed Bi-RNN architectures should be fine-tuned accordingly to meet the required performance. The proposed Bi-RNN channel estimator proceeds as follows

\begin{itemize}
    \item {\ac{ALS}} estimation at the inserted pilot symbols as performed in~\eqref{eq:ALSP}, followed by zero insertion at all the data symbols. Thereafter, the initial estimated channels $\hat{\tilde{\ma{H}}}_{{\rho}} \in \compl^{K_{\text{on}} \times I_{d}}$ are converted to the real-valued domain as performed in~\eqref{eq:lstm_input_stacked}, where $\hat{\Bar{\ma{H}}}_{{in}} \in \mathbb{R}^{2K_{\text{on}} \times I_{d}}$.
    
    \item Bi-RNN end-to-end interpolation, where $\hat{\Bar{\ma{H}}}_{{in}}$ is fed as an input to the optimized Bi-GRU unit. Accordingly, the Bi-GRU unit learns the weights of the estimated channels at the OFDM data symbols. Employing the 2D interpolation using the proposed Bi-GRU unit leads to a considerable performance superiority in comparison with the WI-CNN estimators while recording a significant decrease in the required computational complexity, as shown in Section~\ref{simulation_results}. Also here the proposed Bi-GRU architecture is optimized using the grid search algorithm~{\cite{ref_grid_search}} and trained using the parameters listed in~Table~{\ref{tb:LSTM_params}}. Moreover, similarly as performed in Section~\ref{SBS_RNN_pro}, the performance of Bi-LSTM and Bi-SRNN are investigated in Section~\ref{simulation_results}. 

\end{itemize}

\section{SIMULATION RESULTS} \label{simulation_results}
This section illustrates the performance evaluation of the  {\ac{SoA}} and the proposed RNN and Bi-RNN based channel estimation schemes in terms of \ac{BER} and throughput.
Vehicular communications are considered as a simulation case study, where three mobility scenarios are defined as: (\textit{i}) low mobility ($v = 45~\text{Kmph}, f_{d} = 250$ Hz) (\textit{ii}) High mobility ($v = 100~\text{Kmph}, f_{d} = 500$ Hz) (\textit{iii}) Very high mobility ($v = 200~\text{Kmph}, f_{d} = 1000$ Hz). The power-delay profiles of the employed channel models are provided in Table~{\ref{tb:VCMC}}. It is worth mentioning that in order to guarantee fairness in the conducted simulations, the studied channel estimators are trained using the same parameters shown in Table~{\ref{tb:LSTM_params}}. Moreover, simulation parameters are based on the IEEE 802.11p standard~{\cite{ref4}}, where for the {{\ac{SBS}}} channel estimation, comb-pilot allocation is employed such that $K_{p}I$ pilots are used within the transmitted frame following the comb-pilot allocation. Concerning the FBF channel estimation, the ChannelNet and TS-ChannelNet estimators use $K_{p}I$ pilots per frame, whereas the WI-CNN and the proposed Bi-RNN channel estimators employ only $K_{on}Q$ pilots per frame following the block-pilot allocation, where $K_{on} = 52$ denotes the number of employed subcarriers within the transmitted OFDM symbol, and $Q$ is the number of inserted pilot symbols within the transmitted frame ((i) Low mobility: $Q = 1$, (ii) High mobility: $Q = 2$, (iii) Very high mobility: $Q = 3$). Therefore, the proposed Bi-RNN based channel estimator is able to outperform the recently proposed SoA FBF channel estimators employing fewer pilots with lower computational complexity, resulting in higher transmission data rates as discussed in Section~{\ref{fbf_performance}}.
We also note that these simulations are implemented using QPSK, 16QAM, and 64QAM modulation orders, the SNR range is $[0,5,\dots,40]$ dB. In addition, the performance evaluation is performed according to the employed modulation orders, the mobility scenarios, and variable frame length.

\begin{table*}[t]
\centering
\caption{Characteristics of the employed channel models following Jake's Doppler spectrum.}
\label{tb:VCMC}
\begin{tabular}{|c|c|c|c|c|c|}
\hline
\begin{tabular}{@{}c@{}}\textbf{Channel} \\\textbf{model} \end{tabular}  & \begin{tabular}{@{}c@{}}\textbf{Channel} \\\textbf{taps} \end{tabular} & \begin{tabular}{@{}c@{}}\textbf{Vehicle velocity} \\\textbf{[kmph]} \end{tabular} & \begin{tabular}{@{}c@{}}\textbf{Doppler} \\\textbf{shift [Hz]} \end{tabular}  & \textbf{Average path gains {[}dB{]}}                                                                                       & \textbf{Path delays {[}ns{]}}                                                                            \\ \hline
VTV-UC                 & 12                             & 45                                & 250                                 & \begin{tabular}[c]{@{}c@{}}{[}0, 0, -10, -10, -10, -17.8, -17.8,\\ -17.8, -21.1, -21.1, -26.3, -26.3{]}\end{tabular}     & \begin{tabular}[c]{@{}c@{}}{[}0, 1, 100, 101, 102, 200, 201,\\202, 300, 301, 400, 401{]}\end{tabular} \\ \hline
VTV-SDWW               & 12                             & 100-200                                  & 500-1000                                & \begin{tabular}[c]{@{}c@{}}{[}0, 0, -11.2, -11.2, -19, -21.9, -25.3,\\ -25.3, -24.4, -28, -26.1, -26.1{]}\end{tabular} & \begin{tabular}[c]{@{}c@{}}{[}0, 1, 100, 101, 200, 300, 400,\\401, 500, 600, 700, 701{]}\end{tabular} \\ \hline
\end{tabular}
\end{table*}

\begin{figure*}[h!]
 \centering	\includegraphics[width=2\columnwidth]{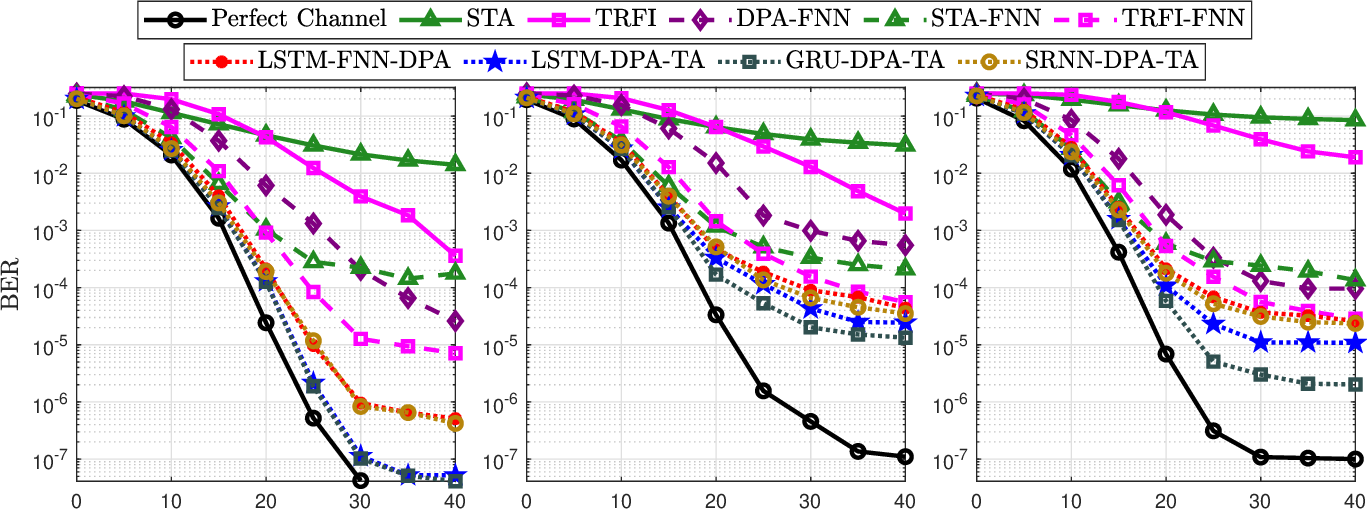}\\[-3ex]
	\subfloat[\label{BER_QPSK} BER performance employing QPSK.]{\hspace{.5\linewidth}} \\[-2ex]
	\vspace*{10pt}
\includegraphics[width=2\columnwidth]{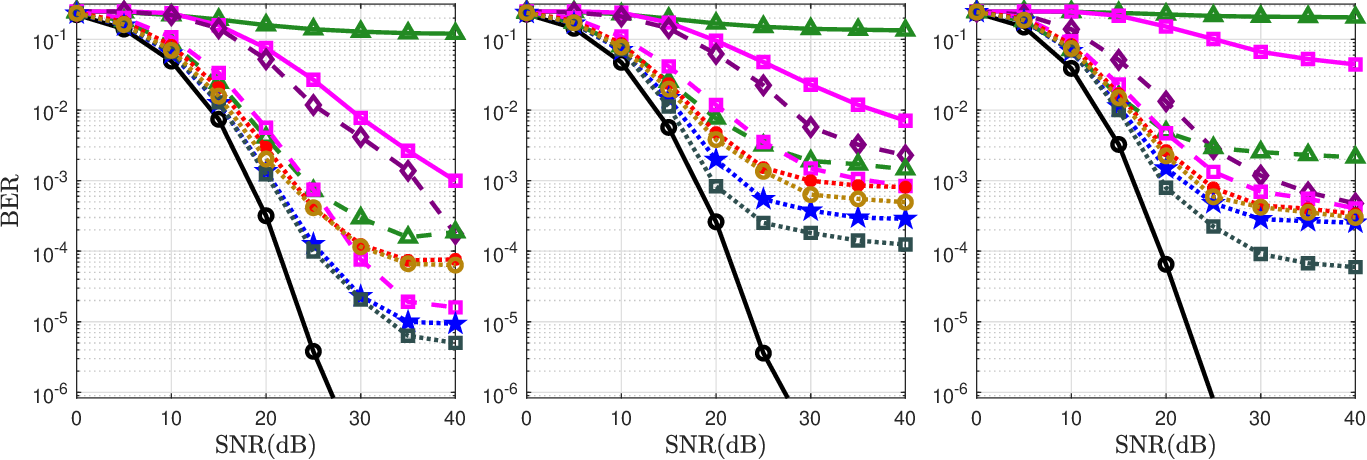}\\[-3ex]
	\subfloat[\label{BER_16QAM} BER performance employing 16QAM.]{\hspace{.5\linewidth}}\\[-2ex]
	\vspace*{10pt}
\includegraphics[width=2\columnwidth]{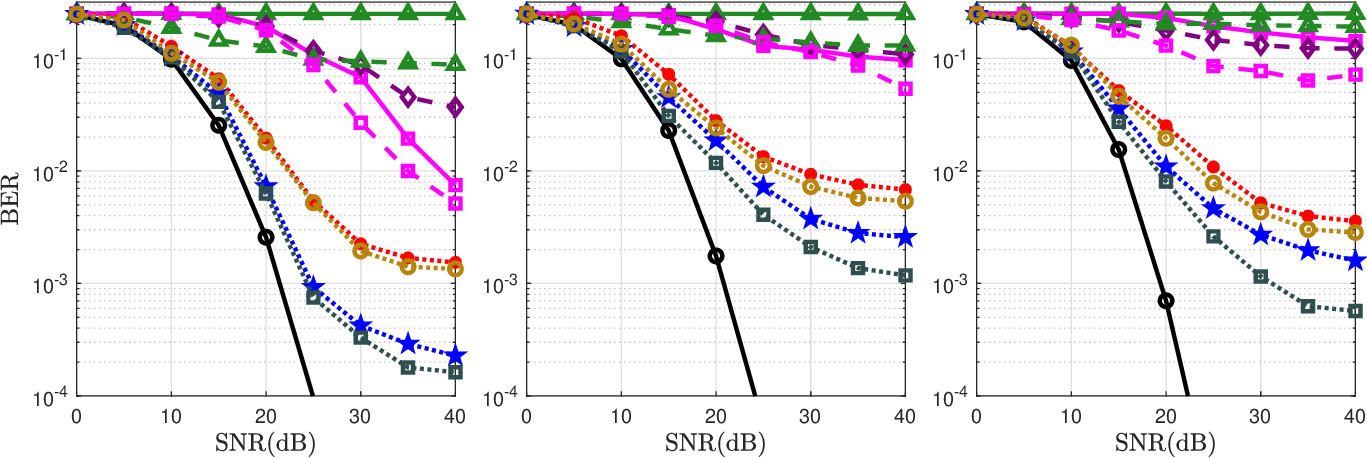}\\[-3ex]
	\subfloat[\label{BER_64QAM} BER performance employing 64QAM.]{\hspace{.5\linewidth}}
	\caption{BER for $I = 100$, mobility from left to right: low, high, very high.}
	\label{fig:BER}
\end{figure*}

\subsection{SBS CHANNEL ESTIMATION}\label{sbs_performance}

\subsubsection{Modulation Order}
For QPSK modulation order, we can notice from \figref{BER_QPSK} that \ac{FNN}-based channel estimators can implicitly learn the channel frequency correlation apart from preventing a high demapping error arising from conventional DPA-based estimation, where STA-FNN and TRFI-FNN outperform conventional STA and TRFI estimators by at least $15$ dB gain in terms of SNR for BER $= 10^{-3}$. However, STA-FNN suffers from an error floor beginning from SNR $= 20$ dB, particularly in very high mobility scenarios. This is due to the STA frequency and time averaging operations that can alleviate the impact of noise and demapping error in low SNR regions. On the other hand, the averaging operations are not useful in high SNR regions since the impact of noise is low, and the STA averaging coefficients are fixed. Therefore, TRFI-FNN is used to improve the performance at high SNRs to compensate for the STA-FNN performance degradation in the high SNR regions. We can clearly observe that employing {\acp{RNN}} as a pre-processing unit rather than a simple FNN in the channel estimation brings a significant improvement in the overall performance. This is because {\acp{RNN}} are capable of efficiently learning the time correlations of the channel by taking the advantage of the previous output apart from the current input in order to estimate the current output. Even though the recently proposed LSTM-based estimators are able to outperform the FNN-based estimator, but using LSTM in the channel estimation is not the best option, due to LSTM long-term memory problem.
In contrast, we can notice that the proposed GRU-DPA-TA estimator is able to outperform the LSTM-DPA-TA estimator by around $6$ dB gain in terms of SNR for BER $= 10^{-5}$, especially, in very high mobility scenario. This is due to the fact that LSTM employs long-term memory, thus, the current estimated channel is affected by older estimated ones. This process harms the performance as the mobility increases, and the channel at successive received {\ac{OFDM}} symbols becomes uncorrelated. 
Whereas, the {\ac{GRU}} uses shorter memory than {\ac{LSTM}}, Thus, leading to the superiority of the proposed GRU-DPA-TA estimator in comparison with the LSTM-DPA-TA estimator. However, we can notice that in low mobility scenario, both LSTM-DPA-TA and GRU-DPA-TA estimators achieve almost similar performance. This is because of the negligible impact of Doppler interference in low mobility scenario, thus, the channels at successive symbols within the received {\ac{OFDM}} frame are highly correlated. So, considering long or short memory while estimating the current channel will not lead to considerable performance degradation. Concerning 16QAM and 64QAM modulation orders, the proposed GRU-DPA-TA estimator outperforms the LSTM-DPA-TA estimator by more than $5$ dB and $7$ dB gains in terms of SNR for BER $= 10^{-4}$ and BER $= 10^{-3}$, respectively, in very high mobility scenarios, as illustrated in~{\figref{BER_16QAM} and \figref{BER_64QAM}. However, it can be noticed that FNN-based channel estimators suffer from severe performance degradation when 64QAM modulation is employed. This is because of the remarkable accumulated DPA demapping error that cannot be eliminated by simple FNN architectures. A nice observation can be noticed from \figref{fig:BER} where employing {\ac{SRNN}} in the channel estimation performs similarly to the LSTM-FNN-DPA estimator in all mobility scenarios.  This reveals that using {\ac{SRNN}} combined with {\ac{TA}} processing  records similar performance as {\ac{LSTM}} combined with {\ac{FNN}}. In other words, the performance degradation caused by the LSTM long-term memory is compensated by the {\ac{FNN}} network in the {\ac{LSTM}}-{\ac{FNN}}-{\ac{DPA}} estimator. However, {\ac{SRNN}} unit can be used instead to eliminate the LSTM long-term memory problem as well as mitigating the noise by simple {\ac{TA}} processing as the case in the {\ac{SRNN}}-{\ac{DPA}}-{\ac{TA}} estimator.

\figref{fig:THR_QPSK} illustrates the throughput of the studied {\ac{SBS}} channel estimators employing QPSK modulation. It can be seen that the proposed RNN-based channel estimators perform higher throughput than conventional and FNN-based channel estimators, especially in low SNR regions. This is due to the accurate channel prediction.

\begin{figure*}[t]
	\setlength{\abovecaptionskip}{3pt plus 3pt minus 2pt}
	\centering
\includegraphics[width=2\columnwidth]{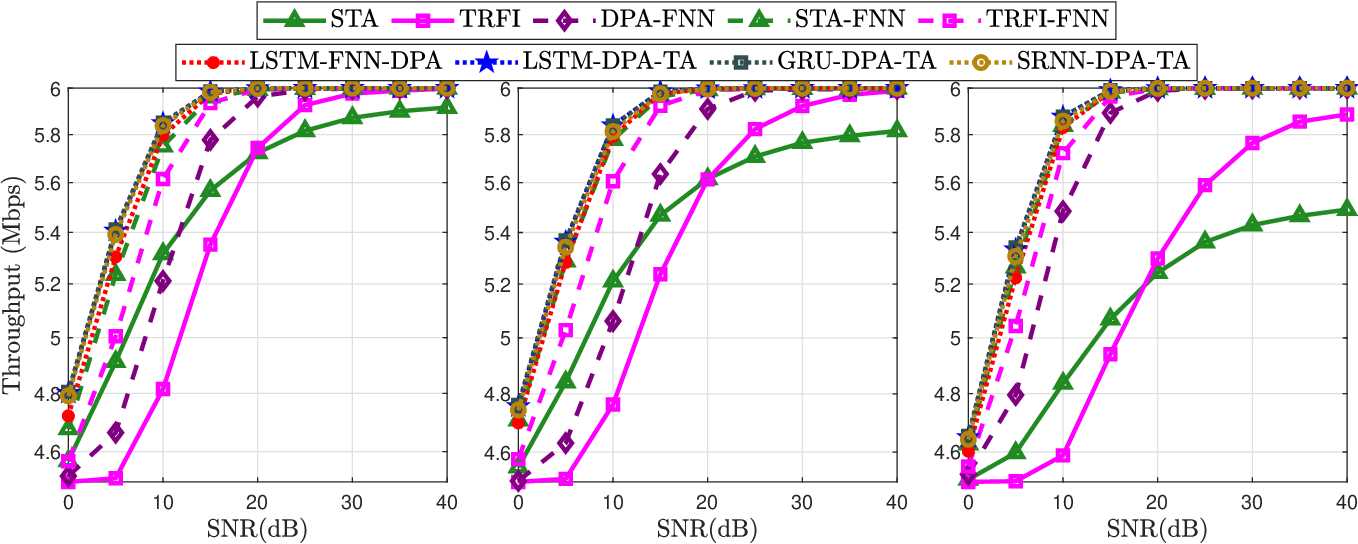} \\
	\caption{Throughput employing QPSK, mobility from left to right: low, high, very high.}\label{fig:THR_QPSK}
\end{figure*}

\begin{figure*}[t]
	\setlength{\abovecaptionskip}{3pt plus 3pt minus 2pt}
	\centering
\includegraphics[width=2\columnwidth]{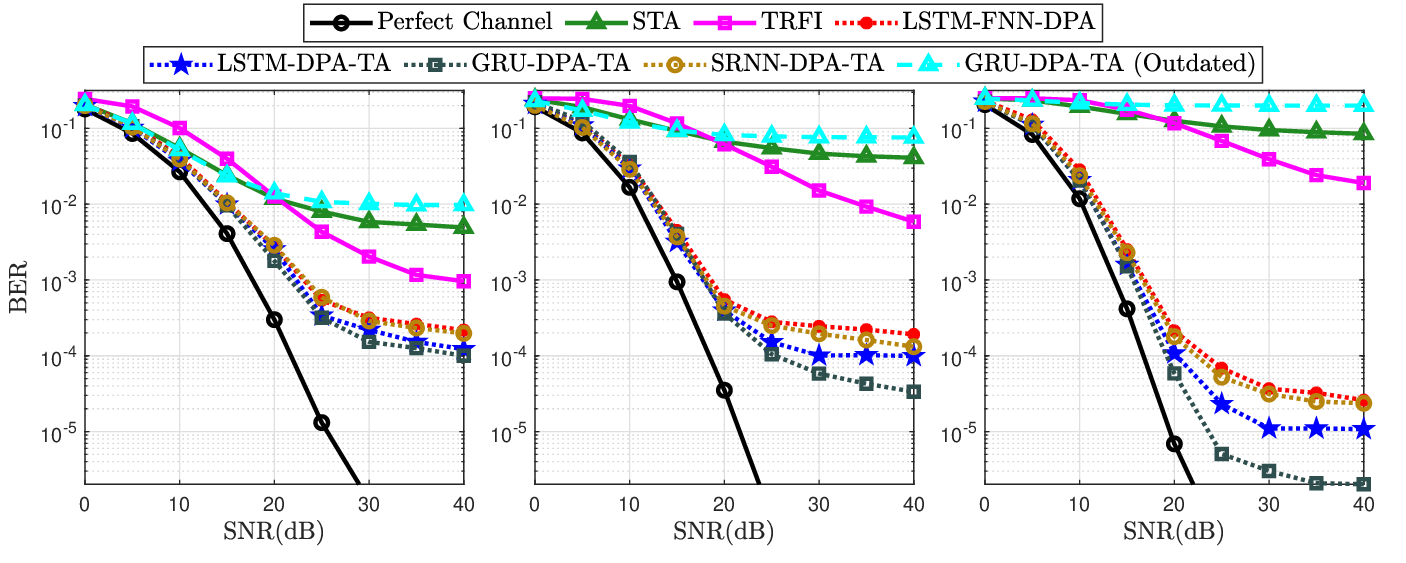} \\
	\caption{BER employing very high mobility and QPSK, frame length from left to right: I = 10, I = 50, I = 100.}\label{fig:FL_QPSK_VH}
\end{figure*}

\begin{figure*}[t] 
    \centering
     \subfloat[Robustness against Doppler frequencies, SNR = 40 dB.]{%
        \includegraphics[width=0.5\textwidth]{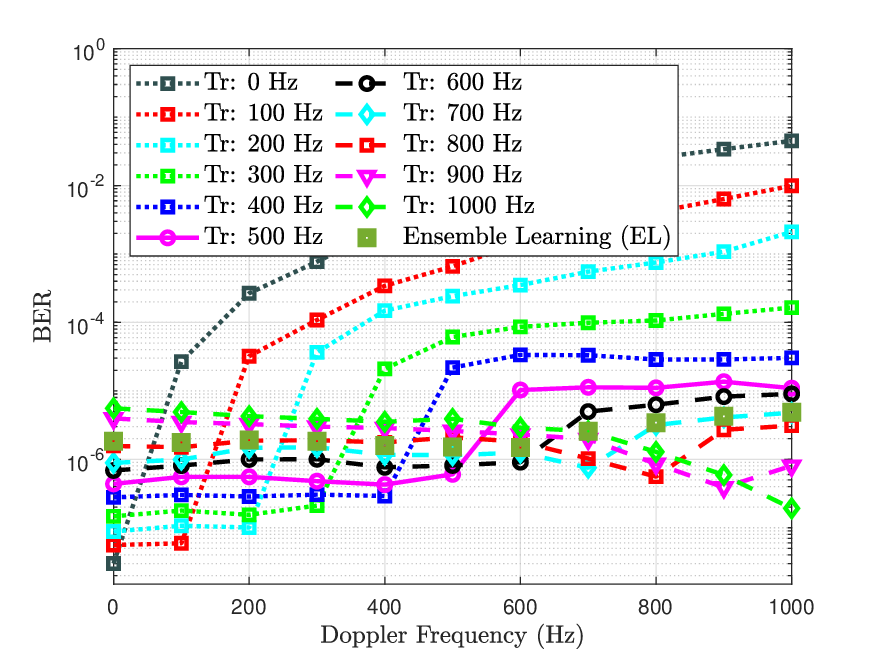}%
        \label{fig:b}%
        }%
    \hfill%
    \subfloat[Robustness against mobility, testing scenario: High mobility.]{%
    \includegraphics[width=0.5\textwidth]{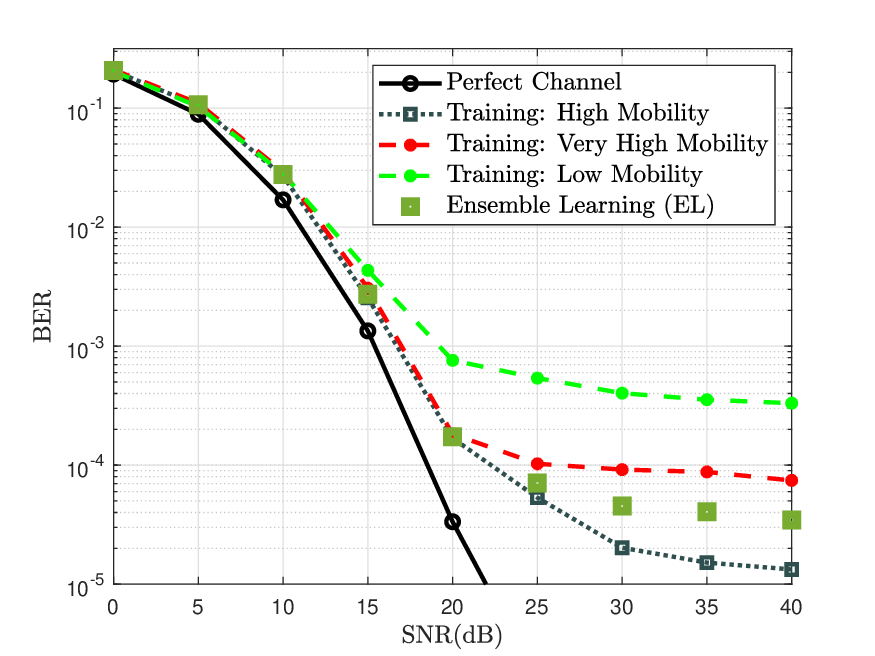}%
    \label{fig:a}%
    }%
    \caption{Robustness analysis of the proposed GRU-DPA-TA channel estimator employing QPSK modulation.}
    \label{fig:rob_mob}
\end{figure*}

\subsubsection{Mobility}

The impact of mobility can be observed in {\figref{fig:BER}}}. The performance  behavior is influenced by the following factors: \textit{(i)} channel estimation error, \textit{(ii)} time diversity due to increased Doppler spread, since the Doppler spread and the time diversity gain are proportional, i.e. more time diversity gain can be obtained in very high mobility scenarios, and  \textit{(iii)} frame length. As shown in \figref{fig:BER}, where the frame length is fixed ($I = 100$),  the performance of all the studied channel estimation schemes degrades with the increase of mobility. This is because the channel estimation error increases with the increase of Doppler frequency. Moreover, we can notice that the conventional STA and TRFI channel estimators suffer from severe performance degradation in the very high mobility scenario, since the impact of the AWGN noise and DPA demapping error is much more dominant than the time diversity gain. On the contrary, the time diversity gain is dominant in DL-based channel estimators, since DL networks are capable of reducing the channel estimation error resulting from the AWGN noise and the DPA demapping error,  leading to a performance improvement in very high mobility scenarios. Note that the net time diversity gain is also related to the employed frame length, since increasing the frame length increases the time diversity gain. This is clearly illustrated in \figref{fig:FL_QPSK_VH}, where QPSK modulation order with very high mobility is utilized. As we can notice, the performance of the proposed RNN-based channel estimators improves when a longer frame length is employed. It is worth mentioning that, the proposed GRU-DPA-TA and LSTM-DPA-TA estimators perform similarly when $I = 10$, since in shorter frames the impact of long and short-term memory cannot be clearly illustrated. On the contrary, when using longer frames, i.e, $I = 50$ and $I = 100$, we can notice the superiority of using the GRU-based estimator instead of the LSTM-based one. 

In order to further illustrate the importance of  channel tracking, \figref{fig:FL_QPSK_VH} shows the performance of the proposed GRU-DPA-TA channel estimator when the outdated estimated channel is used. In this context, the received OFDM symbols are equalized by the DL-based estimated channel at the beginning of the frame. As shown in \figref{fig:FL_QPSK_VH}, equalizing by the outdated estimated channel significantly degrades the performance even when shorter frames are employed. Therefore, this shows the importance of applying channel tracking to guarantee good performance in different mobility scenarios.

Fig.~{\ref{fig:rob_mob}} illustrates the robustness of the proposed GRU-DPA-TA channel estimator against the change in Doppler frequency, where QPSK modulation is employed. Fig.~{\ref{fig:a}} shows the performance of the proposed channel estimator when trained on one Doppler frequency and tested on the entire range of Doppler frequencies. In this context, the entire range of Doppler frequencies is divided into 3 ranges:  (i) Low mobility (0 Hz - 300 Hz), (ii) High mobility (300 - 600 Hz), and (iii) Very high mobility (600 Hz - 1000 Hz). As we can notice, training on one Doppler frequency and testing on the same one gives the best performance. However, training on the highest Doppler frequency within each range shows a satisfactory performance when tested on different Doppler frequencies within the considered range.  This can be explained by the fact that when the model is trained on the worst conditions, i.e., high Doppler frequency, it can perform well when tested on better conditions, i.e., low Doppler frequency. We note that the SNR = 40 dB is considered to show the impact of Doppler interference. Therefore, the performance of the trained models can be generalized, where we can train 3 models only. In addition, we can notice that training on lower Doppler frequencies, (for example, $f_d$ = 250 Hz) and testing on higher Doppler frequencies lead to a severe performance degradation which is expected as shown in Fig.~{\ref{fig:b}}, since the model is trained in the absence of Doppler interference. It is worth mentioning  that, further model generalization can be achieved by using the concept of ensemble learning (EL) in case the velocity range is not known, where the weights of several trained models can be averaged in order to produce one generalized model. We note that, in Fig. 8a, the EL results are obtained by averaging the weights of the trained models on 700 Hz, 800 Hz, and 900 Hz. This combination can be optimized and fine-tuned according to the requirements of real-time applications.

\begin{figure*}[t]
	\setlength{\abovecaptionskip}{3pt plus 3pt minus 2pt}
	\centering
	\includegraphics[width=2\columnwidth]{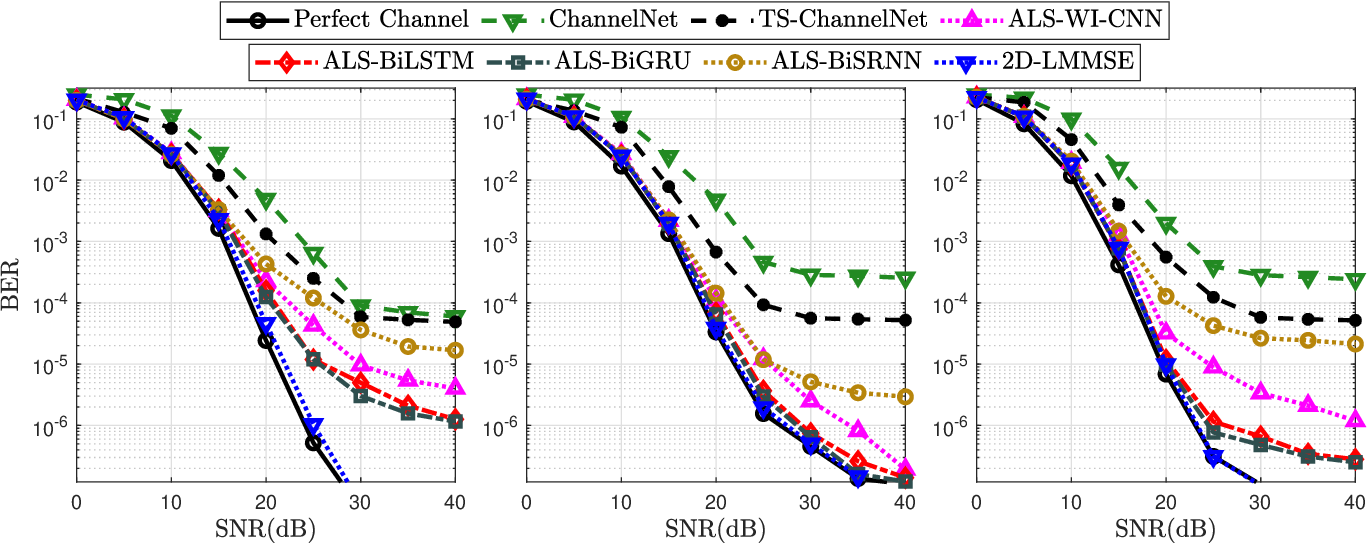}\\[-4ex]
	\subfloat[\label{BER_QPSK_FBF} BER employing QPSK modulation.]{\hspace{.5\linewidth}} \\[-2ex]
	\vspace*{12pt}
	\includegraphics[width=2\columnwidth]{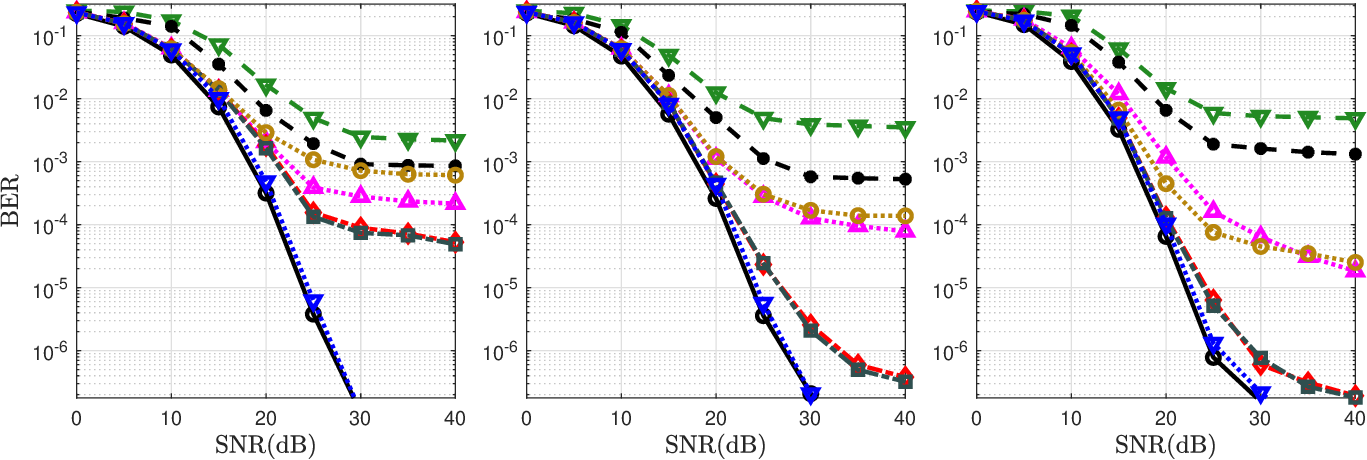} \\[-4ex]
	\subfloat[\label{BER_16QAM_FBF} BER employing 16QAM modulation.]{\hspace{.5\linewidth}} \\
	\caption{BER for $I = 100$, mobility from left to right: low, high, very high. }
\end{figure*}

\subsection{FBF CHANNEL ESTIMATION} \label{fbf_performance}

In this section, performance evaluations of the CNN-based estimators, conventional 2D LMMSE estimator as well as the proposed Bi-RNN based channel estimator are discussed using the same criteria as Section~\ref{sbs_performance}. We note that we only consider the ALS-WI-CNN among the WI-CNN estimators since it has the best performance.

\subsubsection{Modulation Order}

\figref{BER_QPSK_FBF} and \figref{BER_16QAM_FBF} depict the \ac{BER} performance employing QPSK and 16QAM modulation orders, respectively. The performance of {\ac{ChannelNet}} and {\ac{TS-ChannelNet}} accounts of the predefined fixed parameters in the applied interpolation scheme, where the RBF interpolation function and the ADD-TT frequency and time averaging parameters need to be updated in a real-time manner. On the contrary, in the ALS-WI-CNN estimator there are no fixed parameters, and the time correlation between the previous and the future pilot symbols is considered in the {\ac{WI}} interpolation operation. These aspects lead to the performance superiority of the ALS-WI-CNN compared to the ChannelNet and TS-ChannelNet estimators. Although {\ac{CNN}} processing is applied in the {\ac{ChannelNet}}, {\ac{TS-ChannelNet}}, and ALS-WI-CNN estimators, they suffer from a considerable performance degradation that is dominant in very high mobility scenario. This show that the {\ac{CNN}} processing is not able to effectively alleviate the impact of Doppler interference, especially in very high mobility scenarios, where the proposed Bi-RNN based channel estimation scheme outperforms the WI-ALS-CNN estimator by at least $5$ dB and $12$ dB gain in terms of {\ac{SNR}} for a BER = $10^{-5}$ employing QPSK and 16QAM modulations, respectively. We note that the robustness of the proposed Bi-RNN based channel estimator against high mobility is mainly due to the accuracy of the end-to-end 2D interpolation implemented by the utilized Bi-GRU unit. Moreover, we can see that employing Bi-LSTM performs similarly to the ALS-Bi-GRU estimator, this is due to the used frame structure, where the variation of the doubly-selective channel within each sub-frame is low. However, it can be noticed that employing CNN performs better than the Bi-SRCNN unit in low and high mobility scenarios, while using Bi-SRCNN unit leads to around $2$ dB gain in terms of {\ac{SNR}} for a BER = $10^{-4}$ in comparison with the ALS-WI-CNN estimator in very high mobility scenario as shown in~\figref{BER_16QAM_FBF}. As a result, we can conclude that employing Bi-GRU unit instead of {\ac{CNN}} network leads to more accurate channel estimation with lower complexity. Finally, we note that the performance of the 2D-LMMSE estimator is comparable to the performance of the ideal channel but it requires huge complexity as we discuss in the next section, which is impractical in a real scenario. Moreover, the proposed Bi-RNN based estimator records almost close performance as the 2D-LMMSE estimator. Therefore, the proposed Bi-RNN based channel estimator is an alternative to the 2D-LMMSE estimator where it provides a good performance-complexity trade-off.

\subsubsection{Mobility}

The impact of mobility can be clearly observed in \figref{BER_16QAM_FBF}, where the performance of the {\ac{ChannelNet}} and {\ac{TS-ChannelNet}} channel estimation schemes degrades as the mobility increases, and the impact of the time diversity gain is not dominant due to the high estimation error of the 2D RBF and ADD-TT interpolation techniques employed in the {\ac{ChannelNet}} and {\ac{TS-ChannelNet}} estimators, respectively. In contrast, the time diversity gain is dominant in the ALS-WI-CNN and the proposed Bi-RNN based channel estimator, since the initial ALS and WI estimations are accurate, thus,  the SR-CNN and DN-CNN networks are capable of overcoming the Doppler interference. However, using the {\ac{ALS}} estimation at the pilot symbols followed by Bi-GRU unit for 2D interpolation at the data symbols reveal considerable robustness against mobility. This is due to the ability of the optimized Bi-GRU unit to significantly alleviating the impact of Doppler interference, where it can be noticed that the proposed Bi-RNN estimator is able to outperform the ALS-WI-CNN estimators in different mobility scenarios. As a result, the proposed Bi-RNN based channel estimator provides a good performance-complexity trade-off between the CNN-based estimators and 2D-LMMSE estimator.

\begin {figure*}[t]
\centering
\scalebox{0.9}{
\begin{tikzpicture}
\begin{axis}[
    xbar,
    xlabel={Real-Valued Operations},
    symbolic y coords={LSTM-FNN-DPA, LSTM-DPA-TA,GRU-DPA-TA,DPA-FNN,SRNN-DPA-TA,TRFI-FNN, STA-FNN},
    ytick=data,
    xmode=log,
    legend style={at={(0.48,+1.2)},
    anchor=north,legend columns=-1},
    % nodes near coords,
    % enlarge y limits{value=0.2,upper}
    nodes near coords align={horizontal},
    width=2\columnwidth,
    height=8cm,
    grid=major,
    cycle list = {gray!70,blue!80,gray!70,blue!80}
    ]
\addplot+[fill] coordinates {  (133088,LSTM-FNN-DPA) (44136,LSTM-DPA-TA) (22968,GRU-DPA-TA) (10856,DPA-FNN) (10536,SRNN-DPA-TA) (5598,TRFI-FNN) (4810,STA-FNN)};
\addplot+[fill] coordinates { (131432,LSTM-FNN-DPA)  (43488,LSTM-DPA-TA) (22400,GRU-DPA-TA) (10336,DPA-FNN) (10016,SRNN-DPA-TA)  (4598,TRFI-FNN) (4570,STA-FNN)};
\legend{Multiplications/Divisions, Summations/Subtractions}
\end{axis}
\end{tikzpicture}
}
\caption{Computational complexity of the studied {\ac{DL}}-based SBS channel estimators.}
\label{fig:bar_graph_LSTM}
\end{figure*}
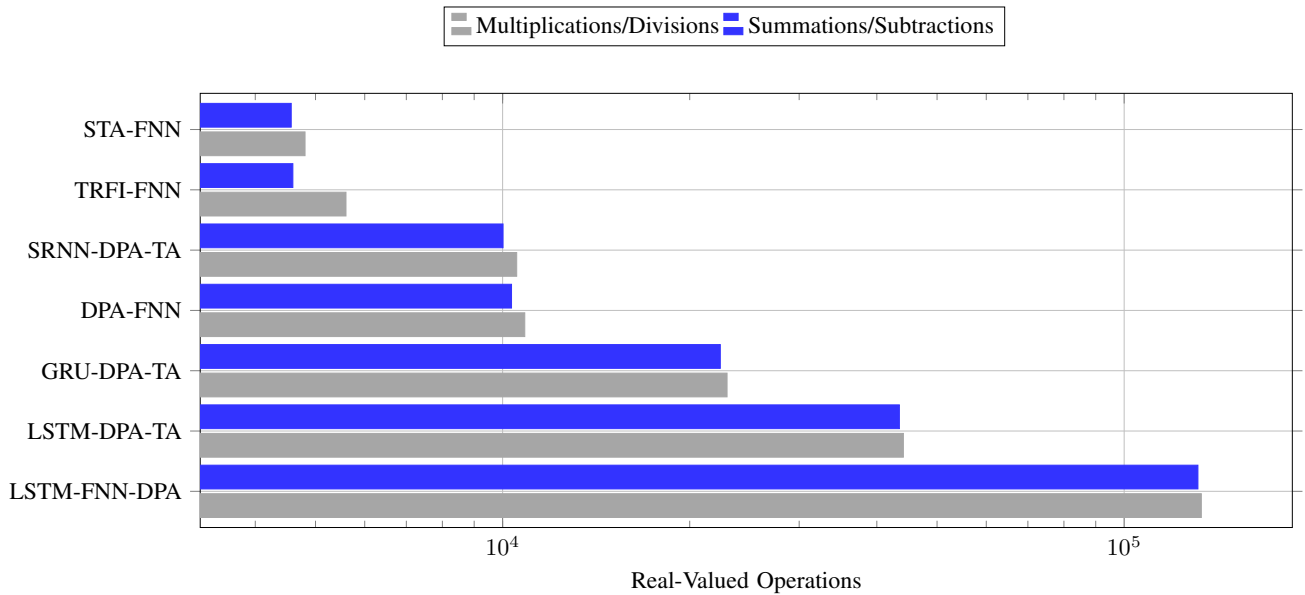

\section{COMPUTATIONAL COMPLEXITY ANALYSIS} \label{complexity}

\begin {figure*}[t]
\centering
\scalebox{0.9}{
\begin{tikzpicture}
\begin{axis}[
    xbar,
    xlabel={Real-Valued Operations},
    symbolic y coords={2D-LMMSE, ChannelNet,TS-ChannelNet,ALS-WI-DNCNN,ALS-WI-SRCNN, ALS-BiLSTM, ALS-BiGRU, ALS-BiSRNN},
    ytick=data,
    xmode=log,
    legend style={at={(0.48,+1.2)},
    anchor=north,legend columns=-1},
    % nodes near coords,
    % enlarge y limits{value=0.2,upper}
    nodes near coords align={horizontal},
    width=2\columnwidth,
    height=9cm,
    grid=major,
    cycle list = {gray!70,blue!80,gray!70,blue!80}
    ]
\addplot+[fill] coordinates {  (3686656161000,2D-LMMSE) (2595149600,ChannelNet) (1180150400,TS-ChannelNet) (428595544,ALS-WI-DNCNN) (36108800,ALS-WI-SRCNN) (2821064,ALS-BiLSTM) (2083008,ALS-BiGRU) (740104,ALS-BiSRNN)};
\addplot+[fill] coordinates { (192000800,2D-LMMSE)  (231045600,ChannelNet) (424060000,TS-ChannelNet) (50263408,ALS-WI-DNCNN) (5789576,ALS-WI-SRCNN)  (2831232,ALS-BiLSTM) (2082944,ALS-BiGRU) (750400,ALS-BiSRNN)};
\legend{Multiplications/Divisions, Summations/Subtractions}
\end{axis}
\end{tikzpicture}
}
\caption{Computational complexity of the studied DL-based FBF channel estimators.}
\label{fig:bar_graph_FBF}
\end{figure*}
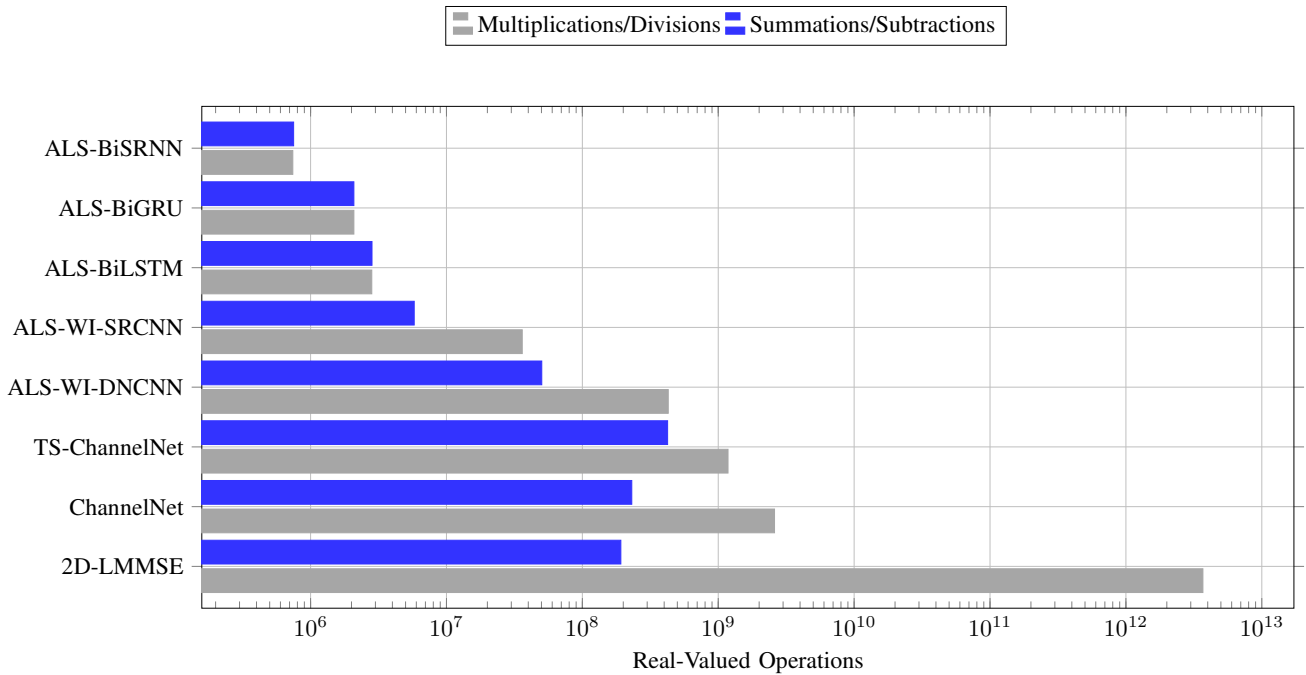

This section provides a detailed computational complexity analysis of the studied channel estimation schemes. The computational complexity analysis is performed in accordance with the number of real-valued multiplications/divisions and summations/subtractions necessary to estimate the channel for one received {\ac{OFDM}} frame~\cite{9813719}.

\subsection{SBS CHANNEL ESTIMATION}

The computational complexity of the  {\ac{SRNN}} lies in the calculation of ${\bar{\ma{h}}}_{t}$ in~{\eqref{eq:RNN_hstate}}, where $P K_{in} + P^{2}$ multiplications and $P K_{in} + P^{2}$ summations are required. The computation of ${{{\ma{o}}}}_{t}$ requires $P^{2}$ multiplications and $P^{2}$ summations. Therefore, {\ac{SRNN}} processing requires $P K_{in} + 2P^{2}$ multiplications and $P K_{in} + 2P^{2}$ summations. Similarly to {\ac{SRNN}}, each LSTM gate requires $P K_{in} + P^{2}$ multiplications and $P K_{in} + P^{2}$ summations. Therefore, the total computations performed by the LSTM unit can be expressed as  $4P K_{in} + 4P^{2}$ multiplications and $4P K_{in} + 4P^{2}$ summations, in addition to $3P$ multiplications and $P$ summations required by~{\eqref{eq: lstm_cell_state}} and~{\eqref{eq: lstm_hidden_state}}. The {\ac{GRU}} unit employs fewer computations compared to the LSTM unit, where it requires $3P K_{in} + 3 P^{2} + 3P$ multiplications and $3P K_{in} + 3 P^{2} + 2P$ summations. 
\figref{fig:bar_graph_LSTM} shows the required multiplications/divisions and summations/subtractions required by various examined {\ac{SBS}} channel estimators. A detailed derivation of the required operations is provided in~\cite{9813719}.

The proposed optimized {\ac{GRU}} unit is configured with $P = k_{d}$ and $K_{in} = 2 K_{on}$. Therefore, it requires $6 K_{on} K_{d} + 3 K^{2}_{d} + 3K_{d}$ multiplications and $6 K_{on} K_{d} + 3 K^{2}_{d} + 2K_{d}$ summations. After that, the proposed GRU-DPA-TA estimator applies the {\ac{DPA}} estimation followed by  the {\ac{TA}} processing that requires $18 K_{on}$ real-valued multiplications/divisions and $8 K_{on}$  summations/subtractions. Hence, the {\ac{GRU}}-{\ac{DPA}}-{\ac{TA}} estimator requires $6 K_{on} K_{d} + 3 K^{2}_{d} + 3K_{d} + 18 K_{on}$  real-valued multiplications/divisions and $6 K_{on} K_{d} + 3 K^{2}_{d} + 2K_{d} + 8 K_{on}$  summations/subtractions. Therefore, the proposed GRU-DPA-TA estimator is able to decrease the required complexity by $48.22 \%$ compared to the LSTM-DPA-TA estimator. In other words, the proposed  {\ac{GRU}}-{\ac{DPA}}-{\ac{TA}} estimator is $2x$ less complex than the {\ac{LSTM}}-{\ac{DPA}}-{\ac{TA}} estimator, at the same time, it achieves a significant performance gain as discussed in Section~\ref{simulation_results}. Employing {\ac{SRNN}} instead of {\ac{GRU}} unit requires $ K_{d} K_{in} +  K^{2}_{d} + 4 K_{d} $ multiplications/divisions and  $ K_{in} + 4 K_{d}  - 2$ summations/subtraction. Therefore it achieves $76.54\%$ and $54.69\%$  computational complexity decrease in the required real-valued operations in comparison to the {\ac{LSTM}}-{\ac{DPA}}-{\ac{TA}} and {\ac{GRU}}-{\ac{DPA}}-{\ac{TA}} estimators, respectively. Therefore, using {\ac{SRNN}} is $4$x and $2$x less complex than using {\ac{LSTM}} and {\ac{GRU}} units. However, the {\ac{RNN}}-{\ac{DPA}}-{\ac{TA}} estimator performs similar to the {\ac{LSTM}}-{\ac{FNN}}-{\ac{DPA}} estimator as shown in Section~\ref{simulation_results}. Finally, we note that a trade-off between the desired performance and the accepted complexity should be taken into account, in order to optimize the use of the {\ac{RNN}}-based channel estimators.

\subsection{FBF CHANNEL ESTIMATION}

The computational complexity of any Bi-RNN unit is twice the required complexity for the regular RNN unit since both forward and backward data flows are applied. The proposed Bi-GRU estimator is optimized where $P = 32$. Moreover, we use $Q = 4$, i.e. assuming very high mobility scenario, in order to have a fair comparison with the ALS-WI-CNN estimator, and $K_{in} = 2 K_{\text{on}} I^{\prime}$, where $I^{\prime} = I + Q$. The {\ac{ALS}} channel estimation at the inserted pilot symbols requires $4 K^{2}_{\text{on}} Q + 2 K_{\text{on}} Q + 2 K_{\text{on}}$ multiplications/divisions and  $5 K^{2}_{\text{on}} Q$ summations/subtractions. Therefore, the overall computational complexity of the proposed Bi-{\ac{GRU}} channel estimation scheme can be expressed by $16 K^{2}_{\text{on}}  + 39946 K_{\text{on}} + 6336$ multiplication/divisions and $20 K^{2}_{\text{on}} + 39936 K_{\text{on}} + 6272$ summations/subtractions. We note that employing Bi-LSTM instead of the GRU unit increases the computational complexity by around $26.29\%$ where  $16 K^{2}_{\text{on}} + 53258 K_{\text{on}} + 8384$
multiplications/divisions and $20 K^{2}_{\text{on}} + 53248 K_{\text{on}} + 8256$  summations/subtractions are needed without any gain in the BER performance as discussed in~\ref{simulation_results}. Moreover, using Bi-SRNN requires $16 K^{2}_{\text{on}} + 13322 K_{\text{on}} + 4096$ multiplications/divisions and $20 K^{2}_{\text{on}} + 13312 K_{\text{on}} + 4096$ summations/subtractions. Therefore, the overall computational complexity is decreased by  $73.63\%$ and $64.22\%$ in comparison to the ALS-BiLSTM and ALS-BiGRU estimators, respectively. However, Bi-SRNN unit suffers from limited performance due to its simple architecture.

\figref{fig:bar_graph_FBF} illustrates the computational complexities of the studied CNN-based FBF channel estimators. We can notice that the conventional 2D LMMSE estimator records the highest computational complexity~\cite{ref_LMMSE_Computational}, making it impractical in real-time scenarios. Moreover, the ChannelNet, {\ac{TS-ChannelNet}}, and the WI-CNN estimators did not provide a good complexity vs. performance trade-off. In contrast, the complexity is significantly decreased by the proposed ALS-BiGRU channel estimator which is $10$x and $115$x less complex than the ALS-WI-SRCNN and the ALS-WI-DNCNN estimators, respectively. Moreover, the proposed ALS-BiGRU channel estimator is $10^{6}$x less complex than the conventional 2D LMMSE channel estimator. Therefore, we can conclude that employing the proposed optimized Bi-GRU network instead of CNN networks in the channel estimation is more feasible and at the same time it offers better performance. Thus making it a good alternative to the 2D LMMSE channel estimator.

\section{CONCLUSION AND FUTURE PERSPECTIVES } \label{conclusions}

In this paper, {\ac{RNN}}-based channel estimation in doubly-selective environments has been investigated. The recently proposed {\ac{DL}}-based {\ac{SBS}} and {\ac{FBF}} channel estimators have been presented and their limitations have been discussed. In order to overcome these limitations, we have proposed optimized {\ac{RNN}}-based and {\ac{Bi}}-{\ac{RNN}} estimators for {\ac{SBS}} and {\ac{FBF}} channel estimation respectively. Moreover, the performance of several {\acp{RNN}} architectures including {\ac{SRNN}}, {\ac{LSTM}}, and {\ac{GRU}} has been thoroughly analyzed based on the channel correlation within the received frame. Moreover, we show that the proposed {\ac{GRU}} and Bi-GRU units result in a better performance-complexity trade-off in different mobility scenarios. Simulation results have
shown the performance superiority of the proposed channel estimators over the recently proposed {\ac{DL}}-based {\ac{SBS}} and {\ac{FBF}} estimators while recording a significant reduction in computational complexity. As a future perspective, we will investigate the ability to extend the proposed RNN and Bi-RNN based channel estimators for MIMO and mmWave communications taking into consideration the constraints of each scenario. In order to further improve the online performance of the proposed channel estimators and their generalization capabilities, advanced DL algorithms such as transfer and meta-learning can be investigated. In addition, more advanced architectures such as the transforms can be tested for channel estimation, considering the trade-off analysis between complexity and performance. Moreover, working on interpretable and explainable theoretical DL models is a crucial future step that would ensure the reliability and transparency of employing DL networks in the domain of wireless communications, especially, channel estimation, where the intuitions behind our proposed work can be further validated.  

\bibliographystyle{IEEEtran}
\small{\bibliography{ref.bib}}

\vfill\pagebreak

\end{document}